\documentclass[preprint, journal]{vgtc}            

\onlineid{1830}

\preprinttext{To appear in IEEE Transactions on Visualization and Computer Graphics.}

\vgtccategory{Research}

\vgtcpapertype{system}

\title{\system: A Dataflow-Based Framework for \\Collaborative Urban Visual Analytics}

\author{%
  Gustavo Moreira, Maryam Hosseini, Carolina Veiga, Lucas Alexandre, Nicola Colaninno,\\Daniel de Oliveira, Nivan Ferreira, Marcos Lage, Fabio Miranda
}

\authorfooter{
  \item Gustavo Moreira and Fabio Miranda are with the University of Illinois Chicago.
  	E-mail: \{gmorei3,fabiom\}@uic.edu.
    \item Maryam Hosseini is with the University of California, Berkeley, and the Massachusetts Institute of Technology.
    E-mail: maryamh@berkeley.edu.
   \item Carolina Veiga is with the University of Illinois Urbana-Champaign.
    E-mail: carolvfs@illinois.edu.
   \item Nicola Colaninno is with the Polytechnic University of Milan.
   E-mail: nicola.colaninno@polimi.it.
   \item Daniel de Oliveira, Lucas Alexandre, and Marcos Lage are with the Universidade Federal Fluminense.
    E-mail: \{danielcmo,mlage\}@ic.uff.br.
   \item Nivan Ferreira is with the Universidade Federal de Pernambuco.
    E-mail: nivan@cin.ufpe.br.
}

\abstract{%
Over the past decade, several urban visual analytics systems and tools have been proposed to tackle a host of challenges faced by cities, in areas as diverse as transportation, weather, and real estate.
Many of these tools have been designed through collaborations with urban experts, aiming to distill intricate urban analysis workflows into interactive visualizations and interfaces.
However, the design, implementation, and practical use of these tools still rely on siloed approaches, resulting in bespoke systems that are difficult to reproduce and extend.
At the design level, these tools undervalue rich data workflows from urban experts, typically treating them only as data providers and evaluators.
At the implementation level, they lack interoperability with other technical frameworks.
At the practical use level, they tend to be narrowly focused on specific fields, inadvertently creating barriers to cross-domain collaboration.
To address these gaps, we present Curio, a framework for collaborative urban visual analytics. Curio uses a dataflow model with multiple abstraction levels (code, grammar, GUI elements) to facilitate collaboration across the design and implementation of visual analytics components. The framework allows experts to intertwine data preprocessing, management, and visualization stages while tracking the provenance of code and visualizations.
In collaboration with urban experts, we evaluate Curio through a diverse set of usage scenarios targeting urban accessibility, urban microclimate, and sunlight access. These scenarios use different types of data and domain methodologies to illustrate Curio's flexibility in tackling pressing societal challenges. Curio is available at \href{https://urbantk.org/curio}{urbantk.org/curio}.

}


\teaser{
  \centering
  \includegraphics[width=1\linewidth]{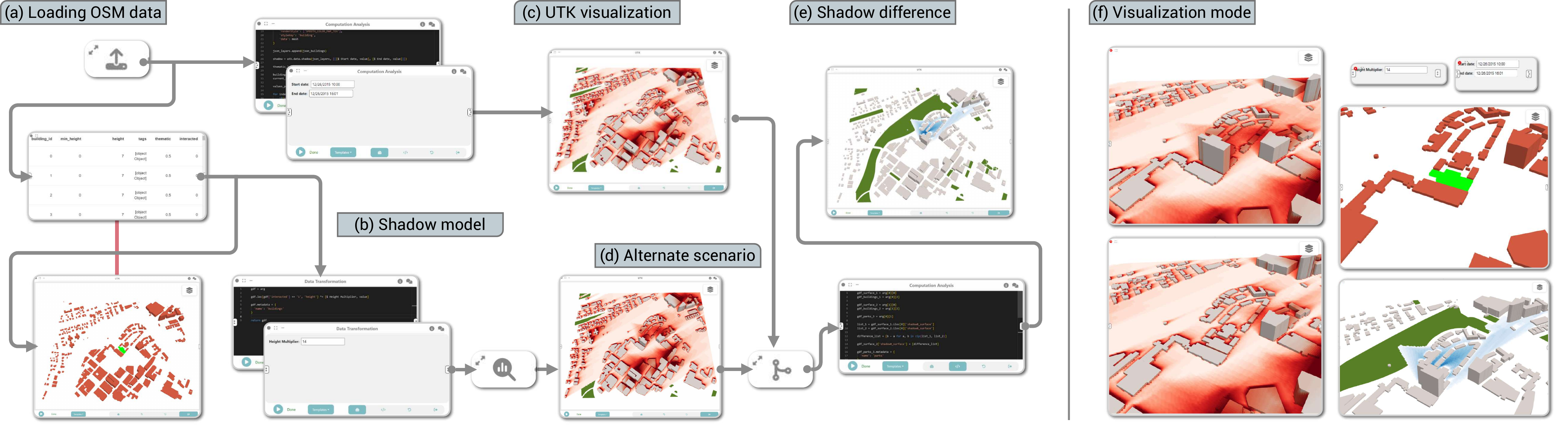}
  \caption{Using \system to build a dataflow to analyze the shadow impact of a proposed building in Boston. (a) The user first adds a node to download OpenStreetMap data for the region of interest, (b) then accumulates shadow data for a period of the day, and (c) visualizes the results in a 3D map. (d) An alternate scenario is created by leveraging \system's interaction node, allowing the user to select buildings in the 3D map. (e) The difference in shadow between the current and alternate states is then visualized. (b,d) With \system, it is also possible to create annotations in the code to facilitate the creation of GUI elements. (f) The final dataflow can be exported as a standalone visualization that includes 3D maps and GUI elements.}
  \label{fig:teaser}
}

\graphicspath{{figs/}{figures/}{pictures/}{images/}{./}} 

\usepackage{tabu}                      
\usepackage{booktabs}                  
\usepackage{lipsum}                    
\usepackage{mwe}                       

\usepackage[hyphens]{url}

\usepackage{mathptmx}                  
\usepackage{amsmath}
\usepackage[dvipsnames,svgnames]{xcolor}
\usepackage{soul}
\setul{1pt}{.4pt}

\newcommand{\Tmrt}{$T_{mrt}$ }
\newcommand{\system}{Curio\xspace}
\newcommand{\hide}[1]{{}}
\renewcommand{\paragraph}[1]{\noindent\textbf{{#1}.}}

\definecolor{blackcolor}{HTML}{000000}
\definecolor{datacolor}{HTML}{545454}
\definecolor{nodecolor}{HTML}{4242a1}
\definecolor{edgecolor}{HTML}{a14242}
\definecolor{modecolor}{HTML}{42a158}

\newcommand{\data}[1]{\setulcolor{datacolor}\ul{#1}\setulcolor{blackcolor}}
\newcommand{\node}[1]{\setulcolor{nodecolor}\ul{#1}\setulcolor{blackcolor}}
\newcommand{\edge}[1]{\setulcolor{edgecolor}\ul{#1}\setulcolor{blackcolor}}
\newcommand{\mode}[1]{\setulcolor{modecolor}\ul{#1}\setulcolor{blackcolor}}

\newcommand{\maryam}[1]{{\color{red} \noindent Maryam: [{#1}]}}
\newcommand{\marcos}[1]{{\color{blue} \noindent Marcos: [{#1}]}}
\newcommand{\fabio}[1]{{\color{purple} \noindent Fabio: [{#1}]}}

\newcommand{\gustavo}[1]{{\color{brown} \noindent Gustavo: [{#1}]}}
\newcommand{\daniel}[1]{{\color{orange} \noindent Daniel: [{#1}]}}

\newcommand{\highlight}[1]{{#1}}

\begin{document}


\firstsection{Introduction}

\maketitle




The growing availability of urban data in the past decade has led urban experts from a diverse range of disciplines to increasingly adopt data-driven methodologies for scientific inquiry and policy formation~\cite{kandt_smart_2021, kontokosta_urban_2021}.
Such data is derived from a multitude of sources, including government databases~\cite{barbosa_structured_2014}, sensor networks~\cite{ang_big_2016}, street-level images~\cite{biljecki_street_2021}, and transportation systems~\cite{zhu_big_2019}. By capturing various aspects of city life, infrastructure, and the environment, data analysis can offer valuable insights to address pressing challenges related to housing~\cite{bin_multi-source_2020}, transportation~\cite{ferreira_visual_2013}, accessibility~\cite{saha_project_2019}, climate~\cite{loibl_effects_2021}, disasters management~\cite{ogie_crowdsourced_2019}, and public health~\cite{luca_crime_2023}.
To effectively analyze urban data, experts must tap into a variety of resources and develop analytics workflows capable of handling the diversity and complexity inherent to urban datasets.
One popular approach is the design of urban visual analytics applications. In collaboration with urban experts, visualization researchers and practitioners construct applications that untangle intricate urban analytics workflows into intuitive, interactive visual representations. This enables experts to explore, understand, and gain insights from data.
Alternatively, urban experts have extensively used computational notebooks (e.g., Jupyter, Observable) to implement their analytics workflows, leveraging a rich ecosystem of urban science libraries~\cite{yap_free_2022}.
Both approaches have their strengths and weaknesses.

On the one hand, urban visual analytics applications tackle research and engineering challenges to enable the interactive visual analysis of data. 
Designing and constructing these applications, however, is a laborious and time-consuming process, requiring visualization researchers and practitioners to account for the inherent complexity of urban data and the diverse needs of experts.
%
Such a complex development process often leads to the creation of one-off monolithic pieces of software that are difficult to extend and adapt to handle new data or use cases.
On the other hand, computational notebooks support literate computing~\cite{knuth_literate_1984} by intertwining code, comments, and visualizations.
However, they offer limited visualization capabilities, being largely constrained to static and simple views that need to be manually specified~\cite{batch_interactive_2018, wongsuphasawat_goals_2019}.
Furthermore, notebooks have been increasingly criticized for encouraging bad programming habits, leading to unexpected execution order, fragmented code, poor versioning, and lack of modularization~\cite{mueller_5_2018, xie_first_2018, grus_i_2018, pimentel_large-scale_2019}.
%

To bridge these gaps and facilitate the creation of urban visual analytics workflows, we propose \system, the \ul{C}ollaborative \underline{Ur}ban \ul{I}nsights \ul{O}bservatory.
%
\system is a web-based visualization framework that allows different users, such as visualization researchers, practitioners, and urban experts, to collaborate in the design and implementation of urban visual analytics workflows.
Such collaboration takes place in a shared interface that uses an intuitive dataflow diagram where users externalize their workflow design decisions through a series of urban-specific computing modules.
Unlike previous dataflow works that were designed for scientific visualization~\cite{haeberli_conman_1988, parker_scirun_1995,bavoil_vistrails_2005} or tabular data~\cite{yu_visflow_2017}, Curio was specifically designed considering common urban data analytics tasks.
Its modules enable users to execute a comprehensive series of tasks seamlessly. These tasks include loading 2D and 3D spatial data that describe the built environment (e.g., street networks, parks, buildings), accessing open-data APIs, cleaning, transforming, and filtering data, as well as creating plot- and map-based interactive visualizations that can be connected through linking and brushing.

\system contains a series of pre-defined, reusable modules and also allows for the creation of user-defined ones. 
These modules can be specified at three abstraction levels: programming using Python, specifying visualizations using Vega-Lite~\cite{satyanarayan_vega-lite_2017} or our previously introduced urban-specific grammar~\cite{moreira_urban_2024}, or through GUI elements.
Following such an approach allows users with different expertise to collaborate and quickly iterate over different design choices, offering functionalities that meet the expectations of both experts and visualization researchers.
Importantly, rather than a compartmentalized development approach that confines visualization researchers and urban experts within the boundaries of their respective domains, \system introduces a common canvas in which they can articulate their intent while maintaining awareness of the contributions and artifacts generated by their collaborators.

Moreover, to ensure the tracking of dataflow modules, \system is a provenance-aware framework that records the evolution of artifacts throughout the collaborative process. It provides a history of modifications, iterations, and contributions made by each user.
In doing so, \system facilitates the exploration of alternative development paths, allowing users to easily revisit and revert changes if necessary.
\system also supports reproducibility through the sharing of complete workflows or specific modules. Users can save and export self-contained descriptions of the workflow and share them with collaborators or other audiences.
\highlight{Curio is available at \href{https://urbantk.org/curio}{urbantk.org/curio}.}
%
%
Our contributions can be summarized as follows:
\begin{itemize}[noitemsep,nolistsep,leftmargin=*]
    \item We present a provenance-aware dataflow that supports urban visual analytics through the combination of modules.
    \item We present Curio, a web-based framework that supports the \highlight{asynchronous} collaborative creation of urban visual analytics dataflows.
    \item We present a series of usage scenarios created in collaboration with urban experts, highlighting Curio's flexibility in addressing several domain problems.
\end{itemize}


%





\hide{
To track and store the different versions of artifacts created in the collaborative process, \system contains a provenance

, grounded on a common set of connected modules.

To allow a more seamless collaboration in the design of urban workflows, \system 

Also, to account for the diversity in users' programming expertise, Curio's modules can be created or modified by the user through low-level programming, high-level grammars, or widgets.

capture experts' workflow decisions


responds to research gaps?

Worse yet, notebook-based approaches have been increasingly criticized for encouriging bad habits that impact reproducibility, such as unexpected execution order with fragmented code, bad versioning and modularization.

offer experts the ability to intertwine code, comments, and visualizations.

collaborative: \url{https://arxiv.org/pdf/1911.00568.pdf}

Though they support literate computing~\cite{?} by intertwining code, comments, and visualizations, and facilitate reproducibility of outcomes, these notebooks offer limited visualization capabilities, primarily constrained to static views.

to facilitate data analysis.

end-to-end urban-specific applications while accounting for data challenges, experts' diverse needs, visualization knowledge gaps, and already-existing workflows.

are custom-tailored 

In contrast to off-the-shelf GIS tools, such as ArcGIS and QGIS, urban visual analytics applications offer

In order to analyze urban data, experts need to tap to different resources and create analytics workflows that are able to account for the diversity and complexity of urban data.

In order to analyze such data, urban experts dispose of different approaches.

create or leverage already-existing analytics workflows.

dispose of a variety of approaches, each with their own strengths and shortcomings.

As this data becomes available and with the .. experts across a widening range of domains use data analysis to tackle problems in X, Y, Z, among others.
In order to better understand and address the multitude of challenges faced by cities today, experts and stakeholders create analytics workflows to gain actionable insight from data~\cite{kontokosta2021urban}.
These workflows can take different shapes, depending on the of specialized expertise.
One popular approach is through the design of urban visual analytics applications. In collaboration with experts, visualization researchers construct applications that untangle intricate urban analysis workflows into intuitive, interactive visual representations, enabling urban experts and stakeholders to explore, understand, and gain insights from data.
Another increasingly popular approach, particularly with 

in the absense of visualization expertise and programming kwnoeldge

In the absense of a visualization researcher, then...

These workflows are often implemented

These workflows often go go beyond simple tabular data and incorporate a wide array of types and sources to better capture the multifaceted nature of urban environments and problems.
Because of 

This complexity requires the adoption of novel approaches for their implementation, such as the creation of urban visual analytics applications or computational notebooks (e.g., Jupyter) leveraging a rich ecosystem of data science libraries.

to address pressing challenges faced by cities today.

of data-driven methodologies for scientific investigation~\cite{?} and policy formulation~\cite{?} among urban experts and stakeholders.

policy formulation~\cite{?} and scientific investigation

led to the increasing adoption of data-centric approaches by urban experts and stakeholders~\cite{?}.
These 

In order to capture the multifaceted nature of urban environments and problems, urban data is diver

The diversity of data and tasks

Given the diversity and hete

Creating data analysis workflows that leverage urban data, however, raises a number of important challenges 

Oftentimes, even going beyond simple tabular data and incorporating a wide array of data types and sources to capture the multifaceted nature of urban environments and problems.

In the past decade, there has been an increase in the number of visual analytics applications designed to leverage urban data to tackle pressing issues faced by cities~\cite{?}.
These applications untangle intricate urban analysis workflows into intuitive, interactive visual representations, enabling urban experts and stakeholders to explore, understand, and derive insights from data. 
Designing and constructing these applications, however, is a laborious and time-consuming process, in which visualization researchers need to account for the inherent complexity of urban data, experts' diverse needs, and already-existing workflows.
Such a complex development process lead to the creation of one-off monolithic pieces of software that lack flexibility to be adapted 

With such a complex development process, it is easy for 

flexibility

curio

serendipity

Though reasons are varied, at its root is the inherent complexity in building end-to-end urban-specific applications while accounting for data challenges, experts' diverse needs, visualization knowledge gaps, and already-existing workflows.

Visualization researchers need to support analytical tasks that require data

incorporate spatiotemporal data that goes beyond simpler tabular data, such as 

one-off tools

diverse and complex data, going beyond simple tabular

support

Oftentimes, even going beyond simple tabular data and incorporating a wide array of data types and sources to capture the multifaceted nature of urban environments and problems.
Designing and constructing these applications, however, is a complex process, given the need to incorporate complex data


Despite their contributions to bridging research gaps identified in the visualization~\cite{?} and domain~\cite{?} literature, the practical adoption of these applications remains limited.\fabio{not sure if this pivoting is needed, or even saying a}

In the absense of readily available tools ...

Similarly to what is happening in other areas, urban experts primarily rely on computational notebooks, such as Jupyter, for their data analysis needs.
Though they support literate computing~\cite{?} by intertwining code, comments, and visualizations, and facilitate reproducibility of outcomes, these notebooks offer limited visualization capabilities, primarily constrained to static views.

The many reasons for such a gap have been highlighted in the visualization literature in a more general data science context~\cite{?}, in the urban context one is particularly salient: the inherent complexity in building end-to-end urban-specific applications.
Given the complexity of the development process, it is expected that visualization researchers will fall into well-known pitfalls leading to the creation of monolithic pieces of software

that lack the required flexibility to be incorporated into experts' workflows or extensibility to serve as starting point for future endeavours.

With a complex development process, it is expected that visualization researchers will fall into development pitfalls that will lead to the creation of black box bespoke applications.

A complex development process leads 

Though reasons are varied, one reason for such a gap between what visualization researchers offer and what urban experts expect, is the fact that building end-to-end urban-specific applications is inherently complex

easy to fall for development pitfalls

Though reasons are varied,

Though reasons are varied, at its root is the inherent complexity in building end-to-end urban-specific applications while accounting for data challenges, experts' diverse needs, visualization knowledge gaps, and already-existing workflows.
Such complex development process leads to the creation of black box bespoke applications where experts are only seen as data providers and evaluators~\cite{?}, rather than invested participants of the design process.

In the absence of readily available frameworks flexible enough to account for urban-specific requirements, experts oftentimes rely on computational notebooks, such as Jupyter, for their data analysis needs. Though popular, these notebooks offer limited visualization capabilities, primarily constrained to static views.
Importantly, without 

To bridge these gaps...

The \emph{status quo} is one where urban experts oftentimes rely on computational notebooks, such as Jupyter, for their data analysis needs, using 

. Though popular, visualization components in these notebooks are primarily static, stiffening

where users create visualizations upfront and interactivity is bounded by the parameters predefined in the code.

Such scenario is particularly common in urban data analysis, where applications usually have to balance the needs of users with diverse background, already-existing practices, and data complexities that are characteristic to urban data.

in spite of many lacking formal programming training.
Experts from different urban disciplines (

visualization researchers construct systems that untangle intricate urban analysis workflows into intuitive, interactive visual representations, enabling users to explore, understand, and derive insights from large and complex data.

Not sure what is the best option:
\begin{itemize}
    \item option 1: start with expert workflow to derive insights from data -> talk about how these are usually hpothesis-driven, use ephemeral computational notebooks, lack interaction and exploratory capabilities -> urban VA bridges this gap, but are hard to build
    \item option 2: start with urban VA, popularity -> allow experts to explore and derive insight from data -> hard to build, monolithic, undervalue domain contributions
    \item option 3: start by talking about the gap between interactive visualization and data science workflows -> point to Batch and Elmqvist's paper on EDA gap
\end{itemize}

\marcos{I think all 3 points are relevant and should be in the intro. I would start with a more general discussion and then go to more specific topics. IMO it seems that 2 is a brother discussion, then 3, and finally 1.}

`` However, at the same time, such flexibility creates challenges in code management, comprehension, and development with notebooks. For example, messes in code may accrue during EDA, and data scientists may lose track of their thought processes. To address these issues, several tools have been proposed — such as Variolite [27], Verdant [28], and Fork It [57] — to support fast versioning and history tracking.''

Any option should cover that it is fundamental to bridge the gap between:
\begin{itemize}
    \item extreme 1: ephemeral computational notebooks, quick and easy to develop, lack interactivity | data analysis driven by scripts
    \item extreme 2: urban VA, hard and complex to develop, might not leverage existing codebase and workflows | data analysis driven by GUI
\end{itemize}

closing visualization knowledge gaps: \url{https://diglib.eg.org/bitstream/handle/10.2312/visgap20201108/035-042.pdf?sequence=1&isAllowed=y} (also highlights opening the black box and early prototyping)

In the last decade, there has been an increase in the number of visual analytics systems proposing to leverage urban data to tackle pressing issues faced by cities~\cite{?}.
By collaborating with urban experts and stakeholders (e.g., urban planners, architects, city officials, climate scientists), visualization researchers construct systems that untangle intricate urban analysis workflows into intuitive, interactive visual representations, enabling users to explore, understand, and derive insights from large and complex data.
In this process, 

gap in EDA and interactive visualization: \url{https://drive.google.com/file/d/1WyT_XpQ-s825PeIvqJ4RHTJghRlBj1rS/view}

popularity of EDA, though computational tools are more popular: \url{https://dl.acm.org/doi/pdf/10.1145/3545995}

The construction of urban visual analytics systems is usually a collaborative process that involves visualization researchers and urban experts.
Such process is 

mention 1: the status quo is one where the vast majority of weather VA contributions have focused on building monolithic and isolated prototypes that are hard to use, combine, and extend

mention 2: Moreover, undue focus is given to the VA end product, rather than to the process that led to it – process which might generate just as useful artifacts as the end product itself [?]; and insights gathered during the design and development process are rarely reported – a situation that recently led for a call for the creation of designing for the graveyard, where visualization components and experimentations (that tend to be more complex) are ultimately stored for later re-use

decompose overly complex VA systems (mention grand challenge)

capture intermediate steps

leverage already-existing workflows and scripting knowledge

provenance: recreating and extending visualization results is just as important as the techniques used and the outcome

capture knowledge externalization and sharing, as well as the ->process<-

moving away from one-size-fits-all VA tools

facilitate design for the graveyard

status quo: stiffens research and application by perpetuating the creation of functionalities in a vacuum and by limiting the practical and real-world use of VA contributions

opening the black box: users are relectunt to apply visualizations if they are not able to translated the process from data to vis. Participatory design approach is needed (and we support that)

jupyter notebook adhere to Knuth's literate programming paradigm, but are static -- impact productivity

consequences: increase reproducibility, transferability, extensibility, trust

\fabio{Fabio.}

\begin{itemize}
    \item Building urban VA systems is complex -- amalgamation of different components.
    \item Low reusability and reproducibility -- only prototypes.
    \item Low transferability
    \item Different degrees of programming proficiency.
    \item Something about provenance from both perspectives (VA and urban)
    \item Mention big challenges in VA: interoperability, creation of components, reproducibility
    \item Mention big challenges in urban domain: transparency and trust in data-driven decision-making
    \item Visual programming can help bridge the gap between these two fields
    \item Contributions: (1) conceptual collaborative space, (2) tool, (3) use cases
\end{itemize}

}
\section{Background}
\label{sec:background}




\highlight{Urban analytics workflows usually require multidisciplinary teams with skills in all stages of the data lifecycle~\cite{zheng_urban_2014, kontokosta_urban_2021,deng_survey_2023}}.
\highlight{For example, urban experts (e.g., architects, urban planners, climate scientists) collect data from various sources or generate new data using simulations. Second, they clean and transform the data, using general~\cite{alam_survey_2022} or urban-specific~\cite{doraiswamy_interactive_2018} data management solutions. Third, they use analytical and modeling approaches~\cite{yap_free_2022} to extract features, identify trends, analyze correlations, etc. Finally, experts create visualizations to gain insights into the data and better understand urban problems or phenomena.
The complex nature of urban analytics presents key challenges that increase the analytical bottleneck for urban experts.}
%
For instance, the heterogeneity and volume of urban data require preprocessing techniques to ensure consistency and usability across the processing pipeline. Urban datasets exhibit considerable variation in format and scale, ranging from street-level imagery to 3D building geometries.
Moreover, the dynamic nature of urban environments means that data is constantly changing, requiring analyses to be updated so that insights do not become outdated before they can be leveraged.
Lastly, urban environments introduce multifaceted relationships among data elements, necessitating multiple iterations and refinements of the data pipeline to handle spatial and temporal dependencies.

To tackle some of these challenges, for more than a decade~\cite{zheng_visual_2016, feng_survey_2022, deng_survey_2023}, researchers across disciplines have been creating end-to-end applications that encompass many of the aforementioned steps and seek to tightly integrate interactive visualizations and analytics capabilities.
%
%
The design of these applications easily leads to rather large codebases. For example, our previous applications~\cite{ferreira_visual_2013,ferreira_urbane_2015,miranda_urban_2017,miranda_urban_2020,rulff_urban_2022,moreira_urban_2024} have approximately 100,000 lines of low-level code that took 6 to 18 months and multiple researchers with different computer science expertise to design and develop, from initial conversations with experts to operational versions.
The iterative process of design and development, often including multiple rounds of feedback and revisions with experts \highlight{from multiple disciplines}, contributes to the expansion of the codebase.
%

With such a lengthy and complex process, development is susceptible to various pitfalls. These may include neglecting best development practices, failing to maintain records of intermediate artifacts, and opting to reinvent the wheel rather than adapting existing codebases (including ones created by urban experts).
Such a scenario stiffens visualization research as well as the creation of practical tools for urban experts.
From a visualization standpoint, given the complexity of these systems, iterating over the design space of urban visualizations is difficult, as the implications of changes need to be carefully weighed against their impact on the existing codebase.
The result is a less fluid design process, rarely considering alternative design choices once a commitment has been made.
%
%
From a practical standpoint, such laborious design and development lead to the creation of bespoke and monolithic pieces of software. 
And these applications are rarely made public. In fact, in a recent survey of open urban planning tools~\cite{yap_free_2022}, only one was originally published by the visualization community~\cite{miranda_urban_2017}, despite the abundance of contributions by the community tackling planning problems.

%
This scenario is not unique to urban visual analytics~\cite{wu_defence_2023}, but the complexities of urban data and tasks make the situation more pronounced in urban contexts.
A framework that tackles the aforementioned challenges answers several recent calls from the visualization and urban research communities.
First, it would support the collaborative creation of urban and geospatial analysis pipelines~\cite{wongsuphasawat_goals_2019,deng_survey_2023,ziegler_need-finding_2023}.
%
Second, it would lower the barriers to the creation of reproducible artifacts that can be easily shared and made public~\cite{ziegler_need-finding_2023}.
Third, it would facilitate the creation of transparent and modular visual analytics modules~\cite{janicke_participatory_2020, wu_grand_2023}.
%
Finally, it would support the creation of provenance-aware pipelines that can be used to recover past states~\cite{alspaugh_futzing_2019, degbelo_fair_2021, saha_visualizing_2022, ziegler_need-finding_2023}, including visualizations discarded in the design process~\cite{akbaba_troubling_2023}.


\hide{
\begin{itemize}
    \item Define what is a dataflow
    \item Challenges of urban dataflow
    \item Highlight problems with urban visual analytics
    \item Big picture
\end{itemize}
}

\section{Related Work}

In this section, we review research related to various aspects of this work. In particular, we review challenges in urban visual analytics, construction tools for urban visual analytics, and collaborative visual analytics systems.

\subsection{Urban visual analytics}
%
%
%

A common usage scenario for urban visual analytics tools is to help untangle the complex dynamics within a city by analyzing its present conditions through visual summaries or mining algorithms~\cite{ferreira_visual_2013,lee_visual_2020,deng_airvis_2020}.
%
Other applications involve analytical tasks that consider the 3D nature of the city and, for this reason, include visualizations of 3D environments~\cite{miranda_star_2024}.
This makes the design and implementation of these tools considerably more difficult due to more complex data management, rendering, and integration between physical and thematic data~\cite{mota_comparison_2023,moreira_urban_2024}.
%
%
%
Urban visual analytics tools also provide critical support for assessing the potential impacts of various changes, ranging from policy to infrastructure and urban development, i.e., \emph{scenario planning}. Such analyses can help evaluate and prepare for different possible future conditions.
To do so, these tools employ modeling techniques to simulate the effect of changes in the city and use visualization to quantitatively and qualitatively evaluate these new scenarios~\cite{ferreira_urbane_2015,miranda_shadow_2019,lyu_if-city_2025}.
%
%
%
%

The examples mentioned above showcase the diversity and complexity of urban visual analytics systems. 
These tools are sophisticated pieces of software designed through collaborations with urban experts. 
Their development encompasses intricate data processing and integration with modeling and simulation techniques, employing advanced visualization designs and interactions.
\system builds on previous efforts, including our own~\cite{ferreira_visual_2013, miranda_urban_2017, miranda_shadow_2019, miranda_urban_2020, moreira_urban_2024}, to facilitate the creation of urban visual analytics applications through urban-specific dataflows. These dataflows can be translated into standalone applications that can be shared and reproduced by urban experts.

\subsection{Construction tools for urban visual analytics}

Urban visual analytics systems rely on several toolkits, frameworks, and authoring tools to implement their visualization requirements~\cite{ferreira_assessing_2024}.
Mei~et~al.~\cite{mei_design_2018} presented a design space and named these tools as construction tools.
In our previous work~\cite{moreira_urban_2024}, we proposed a grammar-based construction tool for urban visual analytics. This tool relies on an urban-specific visualization grammar, requiring users to write JSON specifications.


Computational notebooks offer a highly flexible approach to data analysis, making them a popular environment for urban studies and analyses~\cite{wang_how_2019}.
%
%
%
However, notebooks still carry a number of shortcomings (e.g., unexpected execution order, fragmented code, and poor versioning and modularization~\cite{mueller_5_2018, xie_first_2018, grus_i_2018, pimentel_large-scale_2019}).

%
%
%
%
%

Dataflow visualization frameworks provide a powerful solution to the problems mentioned above, by providing intuitive interfaces to specify not only visualizations but also an entire analytical pipeline.
For example, Bavoil et al.~\cite{bavoil_vistrails_2005} proposed VisTrails, which enabled the specification of visualizations via a dataflow diagram based on VTK~\cite{hanwell_visualization_2015}.
A core component of VisTrails was the use of provenance graphs to record the process of constructing the dataflow.
However, the closeness between VTK classes made the dataflow specifications in Vistrails low-level and hard to create. Also, the interaction between the specified views is limited.
Other examples of visualization dataflow frameworks include ExPlates~\cite{javed_explates_2013} and VisFlow~\cite{yu_visflow_2017}.
However, previous frameworks are limited in the data types and visualizations they support, preventing them from being used for urban analytics workflows that involve modeling, simulation, spatial data, and 2D and 3D map-based visualizations.
\system builds on these experiences and uses a dataflow approach to enable the easier creation and modification of urban analytics pipelines, while supporting the provenance of the users' creations. \system also supports the integration of custom modeling and visualization capabilities.


\subsection{Collaborative visual analytics systems}

Collaboration is essential in projects that leverage data, including the ones targeting urban problems.
In these scenarios, experts collaborate in the process of exploring data to extract insights.
In urban analytics, this generally involves processing, modeling, and visualizing data, as well as predictive analysis for scenario planning.
Therefore, there is a need for effective tools that support this complex collaborative process.
Wang et al.~\cite{wang_how_2019} highlighted that collaboration is desirable, but it is not efficiently facilitated by widely used computational notebooks, such as Jupyter notebooks, for several reasons.
For instance, it is difficult to maintain intermediate products and an understanding of the exploration process.
Another observation is that maintaining the provenance of the development can also help in the collaboration process.

As discussed by Isenberg et al.~\cite{isenberg_collaborative_2011}, providing support for collaboration is neither a trivial task nor usually considered when designing visual analytics systems.
This is especially true in urban visual analytics, with few systems reporting collaboration as one of their design goals, which has been suggested as an interesting research direction~\cite{kunze_visualization_2012,deng_survey_2023}.
Some works are exceptions to this.
%
%
For example, Lukasczyk et al.~\cite{lukasczyk_collaborative_2015} proposed a web-based map tool that enables synchronous and asynchronous collaboration between multiple users.
%
%
Sun et al.~\cite{sun_budi_2021} proposed a 3D collaborative environment for urban design.
All these works, however, have different objectives compared to \system, as they focus on specific urban analytics workflows.
In other words, they do not allow for the construction of analytics pipelines, but rather provide a fixed interface with pre-defined analytics scenarios.
Finally, ArcGIS, a general GIS tool, supports collaboration via shared workspaces and synchronization mechanisms.
\system supports collaboration through various features.
%
For example, Curio's modules have levels of abstraction to support users with different programming backgrounds.
Also, users can add and share comments on the dataflow diagram to facilitate knowledge sharing and the creation of data narratives.
Finally, through provenance, users can explore the different versions of the dataflow produced during the collaboration.
%




\begin{figure*}[t!]
\centering
\includegraphics[width=1\linewidth]{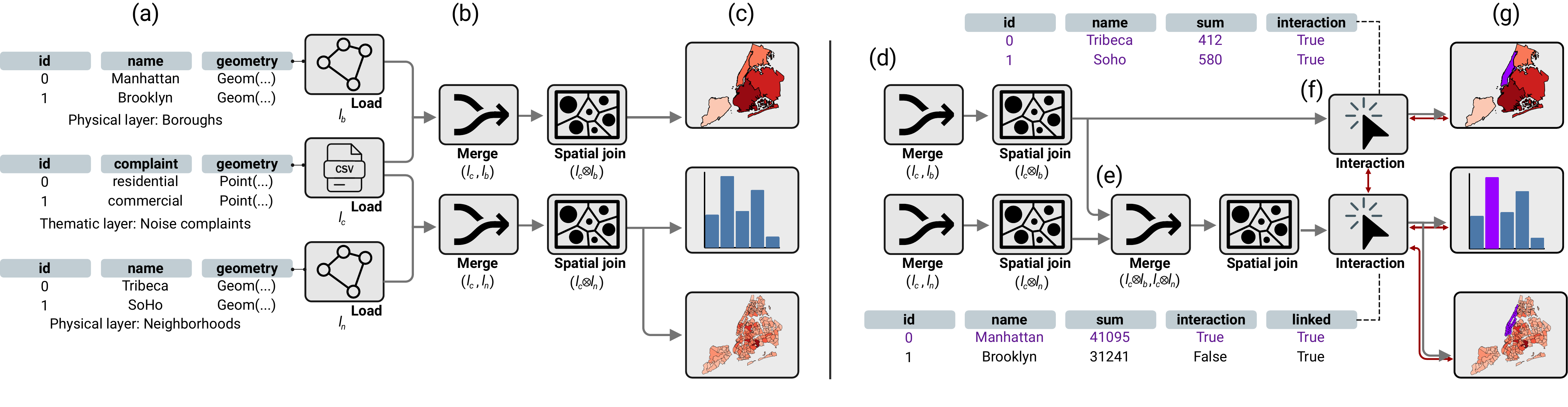}
\vspace{-0.5cm}
\caption{Illustration of the key concepts of the \system dataflow model. (a) A thematic layer $l_c$ and physical layers $l_b$ and $l_n$ are loaded. (b) Spatial joins between $l_c$ and $l_b$ ($l_c\bigotimes l_b$), and $l_c$ and $l_n$ ($l_c\bigotimes l_n$) are computed. (c) The results of the joins are visualized.
(d) To support linked views, the \system dataflow makes use of interaction nodes.
(e) $l_c\bigotimes l_b$ and $l_c\bigotimes l_n$ are further joined, creating a link between them.
(f) Interaction nodes augment the previously joined layers, propagating information when a user selects or brushes elements in a visualization (shown in (g)).}
\vspace{-0.5cm}
\label{fig:interactions}
\end{figure*}

\section{\system's Design Goals}
%
%
\system's overarching goal is to facilitate the flexible creation of urban visual analytics dataflows.
\system's design goals were motivated by our previous contributions (e.g.,~\cite{ferreira_visual_2013,miranda_shadow_2019,miranda_urban_2020,moreira_urban_2024}), as well as meetings with three urban experts (co-authors of this paper).
In these meetings, we were able to better understand their workflows and how visualization tools and computational notebooks were used by them.
This process also shed light on the practical applications of one of our recently proposed toolkits, the Urban Toolkit (UTK)~\cite{moreira_urban_2024}, helping identify key pain points and directly informing our design goals.
Our goals are also influenced by recent works highlighting the need to support collaborative workflows~\cite{wongsuphasawat_goals_2019,deng_survey_2023,ziegler_need-finding_2023}, modular visual analytics components~\cite{janicke_participatory_2020, wu_grand_2023}, and provenance~\cite{alspaugh_futzing_2019, degbelo_fair_2021, saha_visualizing_2022, ziegler_need-finding_2023}.
In the following sections, we refer back to these goals when describing \system.

\paragraph{DG1. Collaborative visual analytics}
\highlight{Urban visual analytics problems are complex and inherently multidisciplinary. For this reason, they are addressed by analysts from different domains (e.g., architecture, engineering, visualization) working in teams.
To enable this, \system should support collaboration in the creation of urban visual analytics dataflows.
The framework should offer features that allow users to contribute to shared dataflows, increasing user engagement in the design process.
Additionally, \system should facilitate an iterative design methodology, encouraging continuous feedback and adjustments.}

\paragraph{DG2. Flexibility and modular design}
As mentioned, urban experts come from diverse disciplinary backgrounds, each with unique methodologies, data requirements, and analytical needs.
To avoid reinventing the wheel with each new urban visual analytics application, \system should support a modular component design.
It should enable users to easily add, remove, or modify modules to meet their specific needs and preferences.
This flexibility extends to the preservation of user-created modules, ensuring their reusability in future projects by allowing users to save and load desired functionalities.
Moreover, recognizing that urban analytics involves experts from diverse fields, \system should support shared component templates.
The framework should also ensure compatibility with a broad spectrum of data formats and standards.

\paragraph{DG3. Reproducibility and provenance of modules}
In a collaborative environment, provenance can enhance efficiency and transparency among collaborators by facilitating users' understanding of the evolution of a dataflow and identifying the sources of data and transformations. Moreover, provenance can also support the collection of design alternatives created in the collaborative process, including ideas that were ultimately set aside in favor of other solutions.
\system should then support the provenance of modules created when users interact with the framework. It should also support the visualization of provenance data and enable users to easily revert to specific versions of modules, facilitating the exploration of their evolution over time.
The framework should also facilitate the reproducibility of modules and dataflows.

\section{A Dataflow for Urban Visual Analytics}

This section introduces the main aspects of \system's dataflow to support the creation of urban-specific workflows and applications, with the aforementioned design principles in mind.
We first formally introduce \system's dataflow model (Section~\ref{sec:dataflow}), followed by the supported types of urban data (Section~\ref{sec:data}).
We then present the dataflow modules (Section~\ref{sec:nodes}) and detail how \system supports data transformation and interactions (Section~\ref{sec:interactions}).

\subsection{The \system dataflow model}
\label{sec:dataflow}

With \system, an urban expert can design and implement workflows by organizing a sequence of steps in a diagram, with steps connected through data dependencies.
These steps compose a dataflow that may include several operations, such as data generation, transformation, management, analysis, and visualization.
In this section, we introduce the main elements of \system's dataflow model: its smallest unit of interest (\data{data layer}), its two classes of computing modules (\node{data node} and \node{interaction node}), and connections (\edge{data dependency} and \edge{interaction dependency}). Our dataflow model formalism extends the one presented by Ikeda et al.~\cite{ikeda_logical_2013}.

In \system, the smallest unit of interest is a \data{data layer} $l$. A data layer is characterized by $k$ attributes (columns) and $m$ records (rows) and can be represented as $l=\{r_1,...,r_m\}$, where each record $r_i$ is a tuple $r_i=(a_{i,1},...,a_{i,k})$, where $a$ are data attributes. \highlight{We denote a set of layers as $L$.}
These data layers can represent a variety of urban data, such as sensor and raster data, images, and building geometries. The different types of urban data supported by \system are discussed in Section~\ref{sec:data}.

A \node{data node} is a module that will represent an operation on top of a data layer, such as data cleaning, transformation, and visualization. These can include pre-defined algorithms or procedures defined by the user. 
A data node $n=(L_i,L_o)$ is then defined as a tuple consisting of input data layers $L_i$ and output data layers $L_o$.
%
%
Each data node takes as an input and produces zero or more data layers, depending on the type of operation. 
\system supports different types of operations so that the framework can be aligned with urban experts' workflows. These steps are detailed in Section~\ref{sec:nodes}.

A \edge{data dependency} $\varphi$ connects the flow of data between two data nodes. It is defined as $\varphi=(L_{flow},n_{source},n_{target})$, with data layers $L_{flow}$ and two data nodes: $n_{source}$ (source node) and $n_{target}$ (target node).
These must satisfy the condition $L_{flow} = n_{source}.L_o \cap n_{target}.L_i$.
The directed dependencies of a dataflow diagram express the connection between two nodes, with the target node having access to the data processed in the source node.

To support user interactions and linked views, \system's dataflow model introduces an interaction node and an interaction dependency.
An \node{interaction node} $n^i$ is a special node that will receive data layer $l$ as an input and will output a data layer $l'$ that is equal to $l$ but with an extra attribute $a_{i,k+1}$ that specifies whether the set of records $R$ were selected or not by the user.
In Figure~\ref{fig:interactions}(d), both tables have an \emph{interaction} column that specifies whether that record was selected by the user.
Furthermore, an \edge{interaction dependency} $\varepsilon$ defines an interaction flow between (1) a data node and an interaction node or (2) two interaction nodes. 
For example, if a dataflow contains two linked plots (defined by two data nodes), each will have data and interaction dependencies to an interaction node; the data dependency will be responsible for passing along data, while the interaction dependency will be responsible for updating $l$ with respect to the records $R$ selected by the user.
It is defined as $\varepsilon= (R,n^i,n)$. An interaction dependency can only exist between two nodes if there is a data dependency between them.
\system's interactions are detailed in Section~\ref{sec:interactions}.

Consequently, a \system dataflow is a composition of multiple nodes, connected by data and interaction dependencies.
Such dataflow can be represented as a tuple $F = (N, N^i, L, \phi)$, comprising data nodes $N$, interaction nodes $N^i$, data layers $L$, and data dependencies $\phi$. 
Interaction dependencies since they only change visualization properties, are not part of $F$. This ensures that $F$ is a directed acyclic graph with respect to the movement of data along the dataflow.
By design, this is a rather flexible dataflow model. As outlined in Section~\ref{sec:nodes}, it supports the creation of urban-specific nodes to handle different parts of a workflow.
%
%
Following such a model, \system supports the collection of historical information regarding dataflow executions for further analysis, i.e., provenance data\cite{herschel1_survey_2017}. 
%
%
%
Our provenance approach is detailed in Section~\ref{sec:provenance}.

\subsection{Urban data layers}
\label{sec:data}

\system supports a variety of urban data layers, covering a wide range of applications and use cases across different urban domains.
Each type of data node, discussed in Section~\ref{sec:nodes}, is restricted to certain input and output layers.
In our previous work~\cite{ferreira_urbane_2015,moreira_urban_2024}, we divided these layers into \emph{thematic} and \emph{physical} data layers.
Thematic data layers correspond to values over 2D or 3D space, such as sociodemographic data over 2D regions or sunlight access values over 3D building surfaces.
Physical data layers correspond to the built or natural environment, such as buildings, road networks, or regions of interest in a city, such as neighborhoods and parks.
In \system, we extend this to also include street-level imagery data, an increasingly popular source of data for urban analyses~\cite{biljecki_street_2021} -- which we call the \emph{image layer}.
Next, we detail these data layers, mentioning key urban datasets as examples.

\paragraph{Thematic layers}
Two types of thematic layers are supported: point and grid layers. These can hold univariate or multivariate data.
A \data{point layer} is used to represent events or features associated with specific locations. Examples include taxi pickups and drop-off positions, environmental observations (air quality monitoring stations, temperature sensors), or positions of noise complaints and crime incidents. Each data point is a data layer record.
A \data{grid layer} (or raster layer) is used to represent data aggregated over a fine-grained grid that covers a region of the city. The grid cells hold data values, such as rainfall volumes or simulated temperatures. Each grid cell is a data layer record.

\paragraph{Physical layers}
Three types of physical layers are supported: 2D and 3D mesh layers, and network layer.
A \data{2D mesh layer} is used to represent geographical boundaries and surfaces in 2D, such as lots, neighborhood areas, water bodies, or any zone that requires delineation over a flat plane. Here, each geographical area (e.g., neighborhood) is a data layer record.
A \data{3D mesh layer} extends the capabilities of the 2D mesh by adding a third dimension to represent the spatial properties of features, such as buildings. In this case, each building is a data layer record.
A \data{network layer} is designed to represent linear features that form networks, such as roads and sidewalks. Here, each network segment (e.g., a street between intersections) is a data layer record.

\paragraph{Image layer}
An \data{image layer} is used to represent street-level image data, such as data from Google Street View or Mapillary. Each entry in this dataset contains, at the very least, a position and an associated image. For images captured by car-mounted sensors (such as Google Street View), this layer can also contain information like the time of capture and whether the camera was facing the left or right side of the street. In an image layer, a tuple with position and image constitutes a data layer record.

\begin{figure*}[t!]
\centering
\includegraphics[width=1\linewidth]{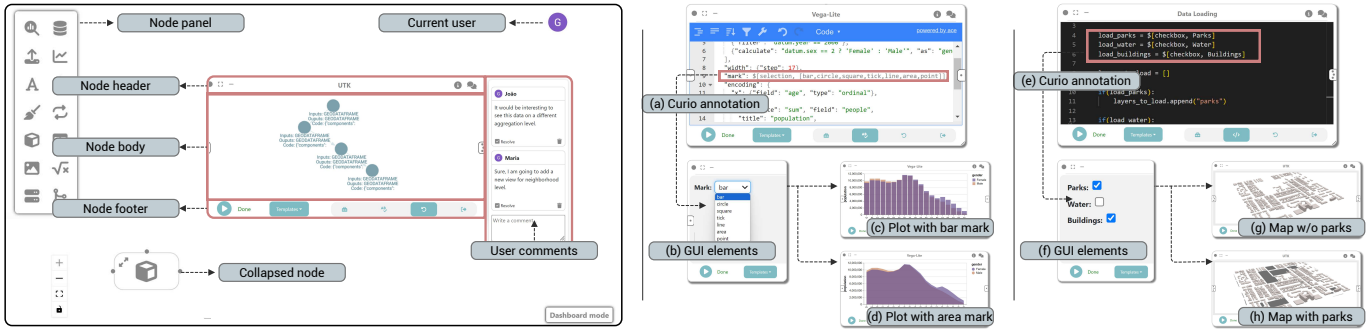}
\caption{Left: Main elements of \system's interface. Center, Right: Connecting facets from the same node. Center: A drop-down menu listing mark attributes is created through an annotation in a Vega-Lite specification. Right: A checkbox is created, but now through an annotation in Python code.}
\vspace{-0.5cm}
\label{fig:annotations}
\end{figure*}

\subsection{Dataflow nodes}
\label{sec:nodes}

\system's dataflow nodes support common data operations performed in urban workflows.
While the framework has pre-defined operations for each one of these nodes, it is reasonable to assume that, due to the diversity of urban workflows, they are not comprehensive.
As such, we present the nodes as \emph{templates} that have defined inputs and outputs, but that can be modified, stored for later use, and shared with other users.
We categorize these nodes into broad categories: data loading, data wrangling \& transformation, analysis \& modeling, visualization, and interaction.
When describing them, we indicate whether they can have zero, one, or more sources and targets in a data dependency.
%
%

\paragraph{Data loading nodes}
\system provides functionalities to load data layers from different sources. 
These nodes are not the target in any data dependency, but they can be the source in one or more.
An \node{OSM node} considers an address or bounding box covering a region of interest and uses that to produce physical layers from OpenStreetMap, including 2D mesh layers, 3D mesh layers, and network layers. 
A \node{raster node} and \node{NetCDF node} load raster data (e.g., from satellite imagery) and NetCDF files (e.g., from WRF simulations).
An \node{open data node} uses the Socrata Open Data API to access common datasets made available by cities.
A generic \node{file node} loads a data file and produces an output with the corresponding layer.

\paragraph{Data wrangling \& transformation nodes}
These nodes are responsible for cleaning and transforming data layers.
They can serve as both the source and target in one or more data dependencies.
For data cleaning, \system provides two basic nodes: the \node{remove duplicates} and \node{remove missing values} nodes.
For data transformation, \system provides the ability to perform \node{statistical normalization}, \node{group-by's}, and \node{spatial joins}.

\paragraph{Analysis \& modeling nodes}
These nodes are available for analytics and simulation tasks.
They can be the source and target in one or more data dependencies.
For modeling, \system provides a node to compute \node{sunlight access}.
This node leverages our previous work~\cite{miranda_shadow_2019} and takes as input one or more physical layers (e.g., buildings, parks) and outputs a 3D mesh layer with corresponding sunlight access values at each vertex.
\system also provides a \node{topology node}, offering connections with the Topology Toolkit~\cite{tierny_topology_2018}, facilitating the extraction of features shown to be useful in urban visual analytics~\cite{miranda_urban_2017}.

\paragraph{Visualization nodes}
\system has four types of visualization nodes.
They can serve as both the source and target in one or more data dependencies.
For a quick overview of the data, we provide a \node{table node} that takes as input one or more thematic layers.
For 2D visualizations, we provide a \node{Vega-Lite node} that also takes as input one or more thematic layers. In this node, the user can directly edit the Vega-Lite specification.
For 3D visualizations, a \node{UTK node} takes as input one or more thematic layers and one or more physical layers, using our previously proposed UTK~\cite{moreira_urban_2024}. Once a data dependency is connected to a UTK node, it automatically creates a basic specification, taking into account the spatial extension of the data given as an input.
For image visualizations, an \node{image node} takes as input one or more image layers. This node will display input images in the form of a mosaic gallery, similar to our previous work~\cite{miranda_urban_2020}.

\paragraph{Interaction node}
This type of node will orchestrate interactions between views. It takes as input one or more data layers, and outputs one or more data layers. Next, this node is discussed in detail.


\subsection{Interactions}
\label{sec:interactions}

Interactions between visualizations are fundamental for enabling deep exploration of intricate collections of urban data.
For this reason, \system's dataflow includes an \node{interaction node} that propagates metadata produced when a user selects or brushes elements in a visualization.
This propagated information can be used to update the visualizations connected to the node.
Figure~\ref{fig:interactions} illustrates an interaction example across a bar chart and multiresolution spatial visualizations (one visualization at a neighborhood level and another one at a borough level).
The user loads a point layer $l_c$ with noise complaints. The layer contains two attributes: position and complaint type.
The user also loads two 2D mesh layers representing areas, $l_{n}$ and $l_{b}$. These layers have two attributes each: area name and 2D mesh of the area.
First, two data transformation nodes perform a spatial join between $l_c$ and $l_{n}$, and $l_c$ and $l_{b}$, using the location of the noise complaints and areas from the 2D mesh layers. 
Then, the data is visualized in three separate views.

In order to support linked views, an interaction node is added. This interaction node adds an extra \emph{interaction attribute} that contains a Boolean value depending on whether a particular data record is selected or not.
The previous visualization nodes are then connected to the interaction node through three new \edge{interaction dependencies}.
It is important to note that this approach selects data records, irrespective of the type of data layer. Given a user selection, the different visualization nodes will be responsible for appropriately selecting the data rows and propagating the information to the interaction node. 
One can also notice an interaction dependency between the two interaction nodes. That connection allows the propagation of interactions between data layers in different resolutions (e.g., neighborhood and borough levels). For this to be possible, an extra attribute must be added to each record of one of the nodes, indicating how that record relates to the records of the other node. \system's templates automatically include that extra attribute.

\hide{
\section{\system Dataflow}
%
This section discusses the main aspects of \system's dataflow. Subsection~\ref{sec:flow_description} explains the graph model that describes a dataflow. Next, Subsection~\ref{sec:data_types} defines the data types supported in a \system's dataflow. Finally, Subsection~\ref{sec:node_types} shows the available data operations.
%
\subsection{Graph Representation}
\label{sec:flow_description}
%
Urban data analysis workflows might involve several steps to derive insight from data~\cite{kontokosta_urban_2021}. These steps compose a dataflow that may include several operations, such as the collection, generation, and discovery of data, followed by its curation and transformation, management, analysis, modeling, and visualization (Figure~\ref{fig:urban_dataflow}).\marcos{add figure: urban dataflow graph} 

As discussed in Section~\ref{sec:background}, a dataflow can be modeled by a directed-acyclic graph (DAG) composed of nodes and directed edges.
The graph nodes represent operations such as data cleaning, transformation, and visualization. Each operation in a dataflow takes as input $0$ to $n_{in}$ datasets. Similarly, each operation produces from $0$ to $n_{out}$ datasets. Nodes may have parameters that modify their operation behavior. The icons in Figure~\ref{fig:urban_dataflow} illustrate the dataflow nodes.
The directed edges of a dataflow graph express the connection between two nodes. If an edge points from node $\sf{N_{a}}$ to node $\sf{N_{b}}$, the dataset outputted by the operation implemented in $\sf{N_{a}}$ is made available as an input of the operation represented by $\sf{N_{b}}$. The gray edges in Figure~\ref{fig:urban_dataflow} illustrate the dataflow edges.

In order to handle user interactions and linked visualizations, we also add the concept of virtual edges to \system's dataflow. 
A virtual edge is an undirected edge that connects nodes, enabling the flow of user-interaction metadata among graph nodes. Since the information transmitted over virtual edges is distinct from the datasets sent through the main dataflow, we assume that these edges do not alter the dataflow graph topology. The red edge in Figure~\ref{fig:urban_dataflow} illustrates a virtual edge.

%
\subsection{Data Types}
\label{sec:data_types}
%
\system dataflows support different types of data. More precisely, it can handle individual values, lists, dictionaries, tabular data, geographic data (both in vector and raster formats), and images.

Individual values store single numeric or string information. It may convey one input parameter of a dataflow operation or hold the output of a computation, such as the percentage of time that a public area is in shadow. 
Indexed collections are represented using lists. An accumulated rain volume dataset collected over time in a weather station can be stored in a list and used as an operation input. Similarly, a list can output rain volume predictions of a numerical simulation performed in a dataflow node.
Key-value pairs are very popular in urban data analysis to represent non-indexed collections and are usually represented using dictionaries. A dictionary can have multiple nested key-value elements, which makes them very flexible to represent complex data.
Tabular data represents a collection of elements and their attributes as a table. Each line of the table represents a data element, and each column of the table represents one attribute. In this way, each cell of the table stores an attribute of an element.
Geographic data can be provided as vector and raster data. Vector data describes geographic elements as a collection of features (points, lines, and polygons) and attributes. It may represent geographical borders, building footprints, streets, points of interest~(POI), \textit{etc.} Raster data represents regions as regular grids. The grid cells hold data values, such as rainfall volumes or elevation.  
Finally, images are arrays of pixels, each storing one color. In urban analytics, images can be used to describe the locations of a city.

\subsection{Urban Analytics Operations}
\label{sec:node_types}
%
\system's dataflow supports all types of data operation usually performed by urban visual analytics systems. 
In fact, it can load data from several file formats, ranging from regular CSV files to complex data formats such as the WRF model~\cite{}. After a dataset is loaded, data-cleaning routines can be employed to ensure the quality of the loaded information.
The dataflow also provides the ability to implement data transformation operations such as statistical normalization and aggregations. Similarly, the \system also foresees the need to perform complex analytic tasks such as topological analysis or machine learning algorithms.
Data processed throughout the \system's flow can be inspected using visualization-specific operations. In fact, exact data values can displayed using the value and table visualization operations. Similarly, images can be displayed using the image visualization operation. 

Abstract data visualization can be produced using grammar-based visualization operations. In fact, statistics charts are some of the most well-known types of visualization. Bar charts, histograms, scatterplots, line charts, and box plots, among others, are mandatory for building effective urban visual analytics applications. Abstract visualizations are not only restricted to statistic charts. Some complex data, such as text and sound, are also visualized using this approach. In fact, texts are commonly represented using word clouds, while sounds are usually shown using spectrograms.

Grammar-based visualization nodes can also produce map visualizations. The visualization of spatial data plays a central role in urban visual analytics systems since data produced by cities are usually associated with geographical locations. This data is oftentimes visualized over a single or multiple maps, which conveys the spatial context of the city. Depending on the urban data characteristics (e.g., spatial dimension) and the tasks performed using the system, both 2D and 3D maps may be used.

The interaction among visualizations is fundamental to enable deep exploration of intricate collections of data. For this reason, \system's dataflow implements a data pool operation that works as an interaction observable that propagates metadata produced when a user picks or brushes elements in a visualization. The propagated information can be used to update the views connected to the pool.  

Finally, the \system's dataflow also allows the definition of a data export operation. This capability is extremely important since it allows us to easily share results produced by a dataflow. Also, it enables the integration of other off-the-shelf tools into dataflow produced in \system.

\marcos{Where should we explain the merge operation}
}
\section{The \system Framework}
\label{sec:framework}
In this section, we present the \system framework, an implementation of the \system dataflow model.
\system is composed of a user interface (detailed in Section~\ref{sec:visual}) and a backend infrastructure (Section~\ref{sec:backend}). 
Figure~\ref{fig:annotations} presents \system's interface and Figure~\ref{fig:architecture} its architecture.
Then, we present the collaborative aspects of the framework (Section~\ref{sec:collaboration}), followed by how it supports provenance (Section~\ref{sec:provenance}), and implementation details (Section~\ref{sec:implementation}).

\subsection{Visual interface}
\label{sec:visual}

\system's visual interface is the central element of the framework. It is divided into two modes: a workspace mode, where users collaboratively create their dataflows; and a visualization mode which streamlines the dataflow into a visual analytics interface.

\subsubsection{Workspace mode}

In the workspace mode, \system provides an infinite canvas in which users assemble their own dataflows by inserting and connecting nodes. 
%
%
This mode is composed of the canvas, the node panel (with the list of available nodes), a visualization mode toggle button, and a user information panel.
When a new dataflow is created, the canvas is initially empty. Nodes can be created by selecting them from the node panel, and they can be resized, repositioned, and deleted.
Once on the canvas, a node is composed of four elements, shown in Figure~\ref{fig:annotations}(left): a header, displaying the type of the node; a body, showing one of four \emph{node facets}; a footer, showing six buttons (run, template selector, GUI mode, programming mode, provenance mode, output mode); and handles on the left and right sides of the node.
Nodes can either be shown in full detail or collapsed into an icon.
An edge can be created by dragging a path between two handles from different nodes. Edges can also be repositioned and deleted.

\paragraph{Node facets}
\system's nodes have four facets. The selected facet is displayed in the body of the node: programming facet, GUI facet, provenance facet, and output facet.
In the \mode{programming facet}, the node's implementation source file is made available to the user and can be freely customized according to their needs. Depending on the node type, either a Python code or visualization grammar specification is displayed and is editable by the user.
In the \mode{GUI facet}, visual interface elements, such as drop-downs, sliders, and checkboxes, are provided to the user, allowing them to adjust parameters and configure the behavior of the node operation.
For example, in an OpenStreetMap data loading operation, the user can define the physical layers of interest (e.g., parks, streets, buildings).
In the \mode{provenance facet}, a tree is displayed with the history of versions of that particular node (Figure~\ref{fig:annotations}(left)). Since nodes can be customized by the user, every time a new version of the node is executed, its version is stored. This approach allows the  user to roll back to previous versions at any point of the dataflow construction.
Finally, the \mode{output facet} displays the data produced by the node, either as a visualization (for visualization nodes) or a formatted output (for all other nodes).

\paragraph{Connecting different facets of the same node}
Though \system comes with a host of pre-defined nodes, as outlined in Section~\ref{sec:nodes}, users can still edit nodes' implementations and make parameters available through GUI elements.
This link between the programming facet and the GUI facet is done through \emph{annotations}. \system's backend interprets the code (either Python or a grammar specification) and searches for specific characters that indicate the beginning and end of an annotation.
An annotation will have the format $\$[type,parameters...]$, where \emph{type} can be one of several GUI elements (checkbox, drop-down, slider, date), and \emph{parameters} are the parameters passed to the element for their construction.
The \system interpreter will translate users' interaction with the GUI elements into Python code and grammar specifications. Once the translation is done, Python code will be sent to the backend, and grammar specifications will be run by Vega-Lite or UTK interpreters.
Figure~\ref{fig:annotations} exemplifies this procedure, for both grammar (a,b,c,d) and Python code (e,f,g,h).
In the grammar example, the user defines a Vega-Lite specification (a) and creates an annotation that will replace the mark property with one of seven marks displayed to the user through a drop-down selection (b). When the user interacts with the GUI element, the node is updated (c,d).
In the Python example, the user modifies a Python code loading OpenStreetMap data and creates an annotation (e) that will replace three Boolean values with the values of three checkboxes (f). Similarly, when the user interacts with the GUI element, the node is updated (g,h).
This design feature allows users to expose desired functionalities to a GUI while retaining the ability to modify and extend the code as needed. 

\paragraph{Node templates}
Each type of node provides users with several pre-defined \emph{node templates}. For example, if the user wants to visualize a scatter plot, they can create a Vega-Lite visualization node and use the scatterplot template. \highlight{Additionally, the framework supports storing and retrieving node templates created or updated by the user. For example, if the user customizes the pre-defined scatterplot template and transforms it into a bubble chart, a new template can be saved in the framework, building an ever-growing template library that can be used by other users in the future.}
These pre-defined node templates can also be updated using annotations to expose common parameters (such as marks and colorscales for Vega-Lite) through GUI elements. \highlight{We note that, since \system's nodes represent common operations performed in urban workflows, node templates enable \system to be adapted for various domain applications and workflows. For example, new data loaders can be added to handle data formats popular in different areas (e.g., OpenStreetMap data, WRF simulations, NASA's SRTM data), and multiple analysis nodes, such as simulation and machine learning techniques, can be added to the system.}
\highlight{In combination with annotations, templates enable users with different backgrounds to collaborate in workflow construction and data analysis tasks by adding new features to the system and receiving feedback, feature requests, and bug reports. For example, these features may help developers understand the needs of domain experts during the implementation of new templates. It also may enable domain experts to create a workflow sketch based on pre-defined node templates and ask for custom features. Additionally, they may allow multiple developers to discuss implementation details or different experts to discuss the workflow steps and results.}

\paragraph{Interactive and linked visualizations}
\system provides visualization nodes for 2D plots, supported by the Vega-Lite node, and 3D urban visualizations, supported by the UTK node.
Both of these nodes support a diverse set of user interactions, such as picking, brushing, panning, and zooming.
For data exploration, \system allows the sharing of interaction states between different visualization nodes, as first outlined in Section~\ref{sec:interactions}. These will be handled by interaction nodes, which are responsible for enriching elements of shared layers with descriptions of the state of the interactions.
In other words, if an element is selected in one visualization node, that information is updated in the interaction node and propagated to all connected visualizations, allowing for interactive updates.

\subsubsection{Visualization mode}
After defining a dataflow, \system provides support for a \emph{visualization mode}, where all nodes and edges are hidden, with the exception of nodes selected by the user to compose a visual analytics interface.
To add a node to the visualization mode, the user can \emph{pin} a node to the interface.
The nodes selected to compose the visual analytics interface can be resized and organized on the screen without losing their original dataflow order and relationships.
This feature is especially important for collaboration. Programmers can use a mix of node templates, Python code, and visualization grammars, and reorganize the elements to present an end-to-end application to urban experts. Figure~\ref{fig:teaser}(f) shows the visualization mode.

\subsection{Backend infrastructure}
\label{sec:backend}
\system's backend infrastructure contains a server, a data manager, and computing sandboxes.
The server is responsible for accepting and responding to HTTP requests from the web-based frontend. It handles user interactions related to the creation and maintenance of dataflows, and orchestrates the communication between the frontend, data manager and computing sandboxes.
The data manager is responsible for storing all datasets loaded and produced during the execution of the dataflow, maintaining the persistence of the dataflow state, and managing the provenance database that tracks of all actions performed during the dataflow construction.
Finally, the computing sandboxes control the execution of the Python code implemented in the dataflow nodes. For security reasons, Python code is executed in a Docker container.

\subsection{Collaboration}
\label{sec:collaboration}

A central aspect of \system is its capability to facilitate the collaborative development of dataflows.
\system offers a user registration feature via a panel located at the top right part of the interface. This enables registered users to collaborate asynchronously on shared dataflows.
To increase collaboration among users with varying levels of programming expertise, \system's nodes have both programming and GUI facets. The GUI facet presents configuration parameters for each node, making it more accessible and user-friendly for users without a programming background.
This strategy can effectively transform a dataflow into a GUI application, similar to tools commonly utilized by urban planners, climate professionals, and social scientists.
Also, \system has a comment feature for nodes (Figure~\ref{fig:annotations}(left)). This allows urban experts and visualization researchers to discuss the addition of new features, the customization of operations, or even the creation of completely new node templates within a dataflow.
This approach documents the collaboration through comments, clarifying the definition and tracking of requirements, and increasing the usefulness of the constructed dataflow.
%
%
Lastly, since \system maintains the version history of all nodes, the user can test different variations of the nodes that were created during the collaboration at any time.

\subsection{Provenance}
\label{sec:provenance}

\system contains a provenance database to track versions of nodes. This database has a schema that encompasses two primary levels: the dataflow level, which includes specifications and executions of the dataflows, and the node level, which details the specifications and executions of individual nodes. Next, we detail the database, highlighting its classes.
At the dataflow level, the database stores information regarding the name of the dataflow and its \ul{nodes}, which consume and produce \ul{layers}. Each \ul{layer} consists of multiple \ul{attributes}. Throughout the collaborative process, several versions of the same dataflow can be generated, for instance, by adding or removing nodes or by modifying the source code of specific nodes.
Consequently, the \ul{user} may execute a transaction (such as adding a node) that results in the creation of a new \ul{version} of the dataflow. Each time a version of the dataflow is executed, a new instance of \ul{dataflow execution} is generated, associated with the specification of a dataflow.
At the node level, nodes' executions are recorded by the \ul{node execution} class, along with instances of layers consumed and produced, represented by the \ul{layer instance} class, and their values stored in the \ul{attribute value} class.
These schema classes can be linked with PROV constructs to generate a provenance graph in the W3C PROV standard~\cite{groth_w3c_2013}. The \ul{user} is mapped to a PROV \emph{agent}; \ul{nodes}, \ul{layers}, and \ul{attributes} to PROV \emph{entities}; and \ul{dataflow execution} and \ul{transformation execution} to PROV \emph{activities}.

\subsection{Implementation}
\label{sec:implementation}

\system's frontend was developed using React.js. React Flow was used to implement the main visual components.
A React component was created for each data node, containing the logic for parsing and displaying the data. Vega-Lite and UTK were used for visualizations.

The backend server was implemented using Python and Flask.
%
SQLite was used to store provenance data, given its simplicity and portability.
Another important part of the backend is the sandbox to run Python code. These containers were configured with the most common libraries used for urban analytics.
Additionally, a new serverless version of UTK was developed, allowing external data from the workflow to be integrated into the framework.
Beyond its visualization capabilities, UTK's Python library was also used to load and parse OpenStreetMap data. To demonstrate the usefulness of these functionalities, we created a set of default template nodes that can be extended or completely redefined by the user.
Finally, user login and registration are managed through Google OAuth.

\begin{figure}[t!]
\centering
\includegraphics[width=1\linewidth]{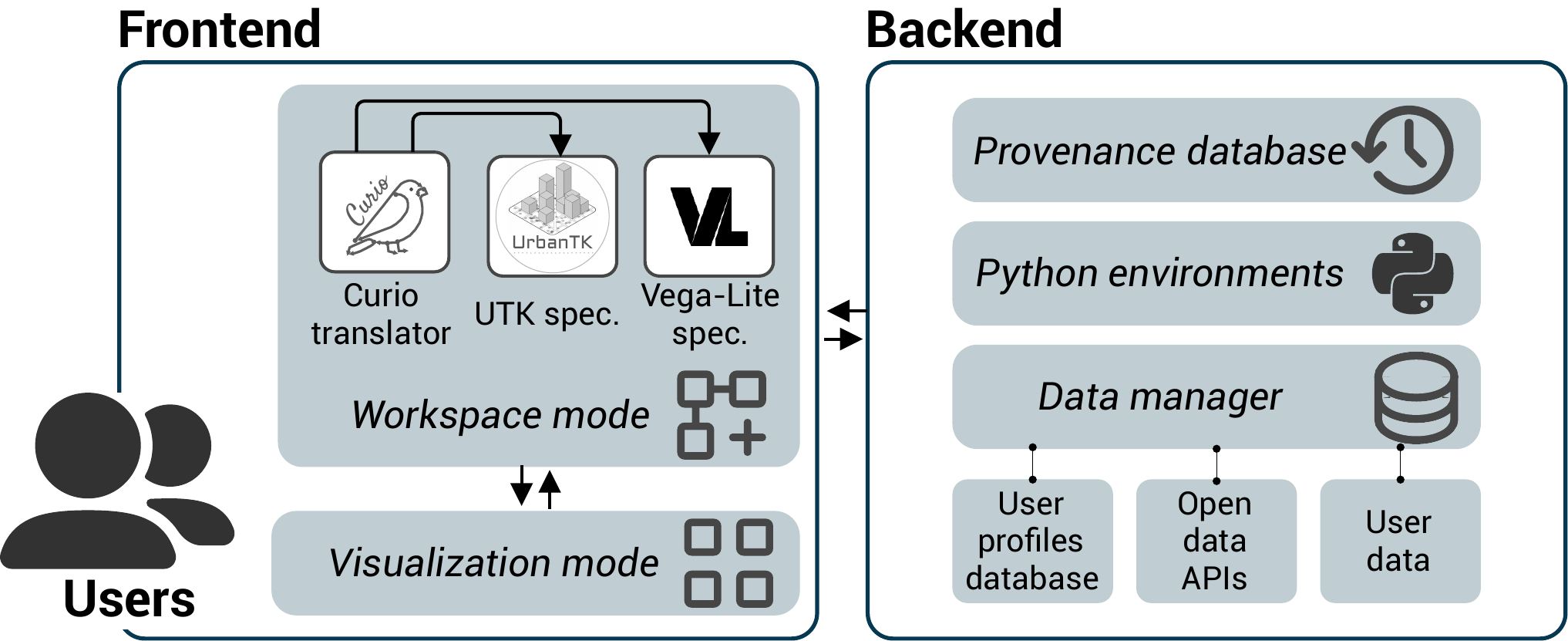}
\caption{\system's frontend and backend components.}
\vspace{-0.5cm}
\label{fig:architecture}
\end{figure}

\hide{
------------

In this section, we describe the \system Framework. As illustrated in Figure~\ref{fig:architecture}, \system is composed of two modules, the User Interface (Section~\ref{sec:interface}) and the Backend Infrastructure (Section~\ref{sec:backend}), detailed in what follows.
%
\subsection{Visual Interface}
\label{sec:interface}
%
The visual user interface of \system is the central element of the framework. It is the visual interface that enables the construction of dataflow, the collaboration of experts and programmers, and the presentation of obtained insights. To achieve these goals, the interface has two modes: the Workspace Mode (Subsection~\ref{sec:workspace}) and the Dashboard Mode (Subsection~\ref{sec:dashboard}), illustrated in Figure~\ref{fig:interface}. Next, we describe the features of \system's visual interface.
 
\subsubsection{Workspace Mode}
\label{sec:workspace}
The workspace mode of \system's visual interface was designed to allow experts and programmers to construct urban visual analytics workflows collaboratively. 
The workspace mode is composed of a canvas, where the dataflow will be constructed and visualized (Figure~\ref{}), the operation menu with the list of available data operations, and a dashboard mode toggle button. 

When a new dataflow is created, the canvas is initially empty to illustrate that no operation has been added to the dataflow. When the user wants to add a new operation, he must use the operations menu to click on one of the icons representing the available operations. Once the desired operation is selected, a box appears in the workspace canvas representing that operation.
If more than one operation is created, multiple boxes will be displayed in the canvas, and the user will be able to create directed edges to connect the operations boxes and compose a dataflow.
Boxes and edges can be deleted if the user no longer desires an operation or a connection between boxes.

\paragraph{Operation Boxes Modes} The boxes used to represent dataflow operations have three usage modes: the expert mode, the developer mode, and the output mode. 
The expert mode (Figure~\ref{fig:interface}) was designed to be used by urban stakeholders without advanced computer science or programming backgrounds, such as urban planners and architects. In the expert mode, visual interface elements such as drop-downs, sliders, text fields, \textit{etc.} are provided to the users, allowing them to adjust parameters and configure the behavior of the operation. For example, in an OpenStreetMap data loading operation, the user may be allowed to define the physical layers (parks, streets, buildings, \textit{etc.}) as well as the bounding box of the area of interest.
The developer mode (Figure~\ref{fig:interface}) was designed to be used by professionals with programming skills. In this mode, the operation's implementation source file is made available to the user and can be freely customized according to their needs. Depending on the box type, either a Python code or a visualization grammar JSON specification is displayed and can be edited by the user.
It is important to highlight that the visual interface elements displayed in the expert mode are defined using annotations added in the Python source or in the visualization grammar specification. This approach allows the creation of custom versions of the operation boxes adequate to the expertise of the specialist. \marcos{Add a good example to illustrate this feature.}
Finally, the output mode displays a chunk of the data produced by the box (in the case of data manipulation boxes) or visual data representations (in the case of visualization boxes).

\paragraph{Operation Templates}
%
Each type of operation box provides users with several predefined operation templates. For example, if the user wants to visualize a scatter plot based on two columns of a tabular dataset, he can create a grammar visualization box and load the scatter plot template. If the user is an expert, he will be able to select the columns of interest using the box expert mode. On the other hand, if the user is a developer, he can edit the visualization grammar specification of the default scatter plot template and customize the chart. 
The use of templates makes \system easy to extend and an analytic tool suitable to build dataflows to handle a diverse range of urban challenges. 

\paragraph{Operation Versioning}
%
Since operations can be customized by the users, either by changing parameters in the expert mode or editing the source implementation in the developer mode, every time a new version of the operation is created and executed, its previous version is stored. Using this approach, the user can roll back to previous versions of the operations at any time of the dataflow construction.
The version history of each operation box can be displayed in the box output mode. The box version history is represented using an interactive graph where each node represents a version of the box operation. In order to navigate among versions, the user needs to select the node representing the version of interest.

\paragraph{Interactive and Linked Visualizations} 
%
\system was designed to enable the construction of urban visual analytics dataflows. For this reason, it is mandatory to have responsive and interactive visualization boxes available during the dataflow creation.
In fact, the system provides several visualization boxes that can be used in any location of the flow. Moreover, visualization grammar boxes that interpret Vega-lite~\cite{} and UTK~\cite{} specifications are also available.
Using these two types of boxes, the user can create complex visualizations ranging from classic statistics charts to complex 3D maps. Also, Vega-lite and UTK allow the implementation of a diverse set of user interactions like picking, brushing, panning, and zooming.
To make the visual data flow even more interactive, it is possible to share data among different visualization boxes by using what we call a data pool box and creating linked visualizations. 
To achieve that, data pools enrich each element of the shared dataset with interaction descriptions. When an element is selected in one visualization box, this information is updated in the data pool and propagated to all connected visualizations, allowing interactive updates.

\paragraph{Collaborative Dataflow Construction}
%
One of the main characteristics of \system is the possibility of urban experts and computer scientists collaboratively creating visual urban dataflows. 
Several functionalities were implemented to make this collaboration effective. 
As discussed earlier in this section, each box has an expert and a developer operation mode. The expert mode exposes the main configuration parameters of each operation using a visual interface approach that would be familiar to anyone without a programming background. 
This approach makes the dataflow construction process similar to previous visual analytics systems used by urban planners, weather professionals, social scientists, \textit{etc}. 
However, since \system's operations can be implemented from scratch and existing templates fully customized by programmers in the developer mode, we implemented a comment tool inside the operation boxes so experts and developers can discuss the addition of new parameters in the expert mode, the customization of the related operation or even the creation of a completely new box template.
Using this approach, the collaboration is documented in the comment tool, making the definition of requisites clearer and increasing the usefulness of the constructed analytics dataflow.
Also, since the system keeps the version history of all boxes, the expert can test the different variations of the boxes produced during the collaboration.  

\subsubsection{Visualization Mode}
\label{sec:visualization}
%
After the definition of a dataflow, the user can change \system's operation mode to the Dashboard Mode. This functionality keeps only the boxes selected by the user on the screen to compose an analytics dashboard. To add a box to the dashboard, the user must click on the icon in the bottom left corner of the box. 
To be precise, in the dashboard mode, all edges and auxiliary dataflow boxes can be hidden to make the interface cleaner and less destructive, facilitating the analytics process.
The boxes selected to compose the dashboard can be freely resized and organized on the screen without losing their original dataflow order and relationships. Figure~\ref{fig:interface} shows an example of a dashboard constructed using \system. \marcos{Is there anything else we should say regarding the dashboard mode?} \gustavo{We could mention that this is a feature for collaboration. Programmers can prepare an interface for experts, constraining less experienced users to an even higher level of experience.} \gustavo{also we have a fullscreen mode for the boxes that hide all controls}
\gustavo{I think it is also important to emphasize how flexible it is to build a dashboard using a mix of templates, low-level code, powerful visualization libraries, and free reorganization of elements visually and logically linked on the workflow mode. So I think this dashboard mode is closer to a VA system than to a dashboard}

\subsection{Backend Infrastructure}
\label{sec:backend}
%
\system backend infrastructure contains three components: The Server Core, the Data Manager, and the Computation Sandbox (see Figure~\ref{fig:architecture}). 
The server's Core Component, implemented using the Python's package Flask, is responsible for serving the \system's interface and accepting its HTTP requests. 
The Core Component can handle several request classes. 
First, it directly handles user interactions related to the creation and maintenance of dataflows. 
Second, it manages the Data Manager Component, which is responsible for the storage of all datasets loaded and produced during the execution of the workflow, for the persistence of the workflow state, and the maintenance of a provenance database that keeps track of all actions performed during the dataflow construction (see details in what follows).
Finally, the Core Component controls the execution of the Python codes implemented in the dataflow activities. The codes are run in the Sandbox Component, which is implemented as an isolated container using docker for security reasons since it executes code provided by the users.

\paragraph{Provenance Database}
%

\daniel{The \system 's provenance database follows the schema outlined in Figure \ref{fig:dbmodel}. This schema comprises 12 classes organized with consideration to both the dataflow space, consisting of the dataflow specification and execution (classes in white), and the version space, which represents the organization of dataflow versions (classes in light grey). The class \textit{Dataflow} stores information regarding dataflows, such as their names, and is comprised of several \textit{Transformations}, which consume and produce \textit{Datasets}. Each \textit{Dataset} consists of multiple \textit{Attributes}. Throughout the collaborative process, several versions of the same dataflow can be generated, for instance, by adding or removing transformations or by modifying the source code of specific transformations. Consequently, a \textit{User} may execute a \textit{Transaction} (such as adding a transformation) that results in the creation of a new \textit{Version} of the dataflow, represented by the \textit{VersionedElement} class. Each time a version of a dataflow is executed, a new instance of \textit{DataflowExecution} is generated, associated with the specification of a \textit{Dataflow}. Transformations' executions are recorded by the \textit{TransformationExecution} class, along with instances of datasets consumed and produced, represented by the \textit{DatasetInstance} class, and their values stored in the \textit{AttributeValue} class. These schema classes can be correlated with PROV constructs to generate a provenance graph in the W3C PROV standard \cite{groth2013prov}. The \textit{User} is mapped to a PROV \textit{Agent}, \textit{Datasets} and \textit{Attributes} to PROV \textit{Entities}, and \textit{DataflowExecution} and \textit{TransformationExecution} to PROV \textit{Activities}.}

\begin{figure}[ht!]
\centering
\includegraphics[width=.9\linewidth]{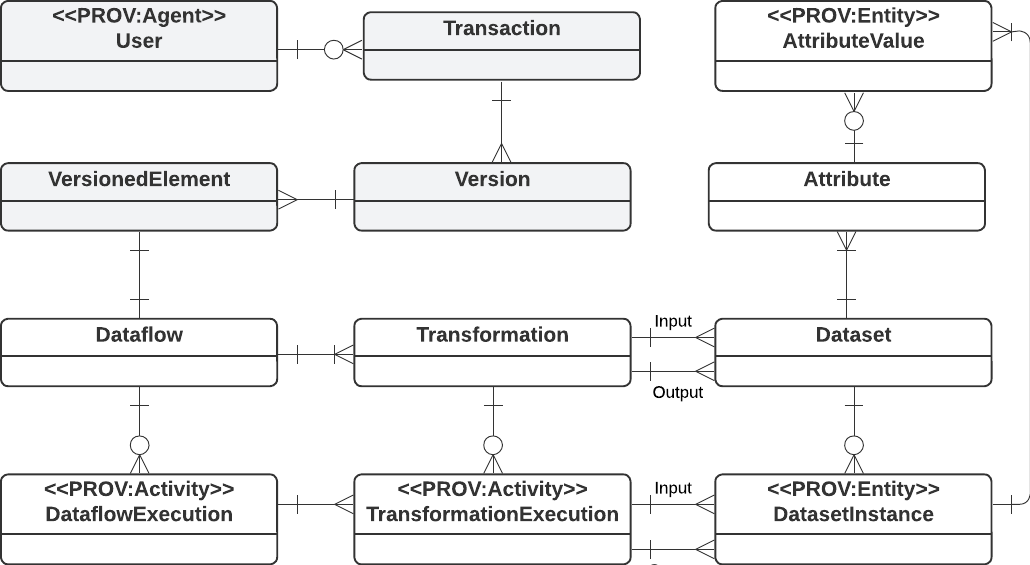}
\caption{Provenance database schema.}
\label{fig:dbmodel}
\end{figure}

}
\section{Usage Scenarios}
\label{sec:use}

In this section, we present a set of usage scenarios that demonstrate \system's flexibility in creating dataflows to tackle pressing urban issues.
\highlight{The scenarios were created in collaboration with three urban experts: two urban planners with experience in urban accessibility (E1) and urban microclimate (E2), and one climate scientist (E3).} All of them hold PhDs.
First, we engaged with them in a series of one-hour interviews during which we asked them to present some of their common workflows, from data collection to visualization.
We then collaboratively built a series of dataflows using \system. \highlight{In this process, the urban experts were responsible for creating data analysis and modeling nodes, while visualization researchers handled data wrangling and transformation, and visualization nodes. This collaboration occurred asynchronously, using Curio's comment feature to track changes and updates.}
The scenarios tackle distinct challenges by integrating a diverse array of data, including images, 3D models of buildings, weather simulations, and sociodemographic data, drawn from various urban settings, such as Boston, Chicago, and Milan.
The cases highlight the steps of the dataflow construction process.
We direct the reader to the supplementary video for an overview of \system in action, \highlight{as well as to \system's webpage for step-by-step overviews of the scenarios.}


\begin{figure*}[ht!]
\centering
\includegraphics[width=1\linewidth]{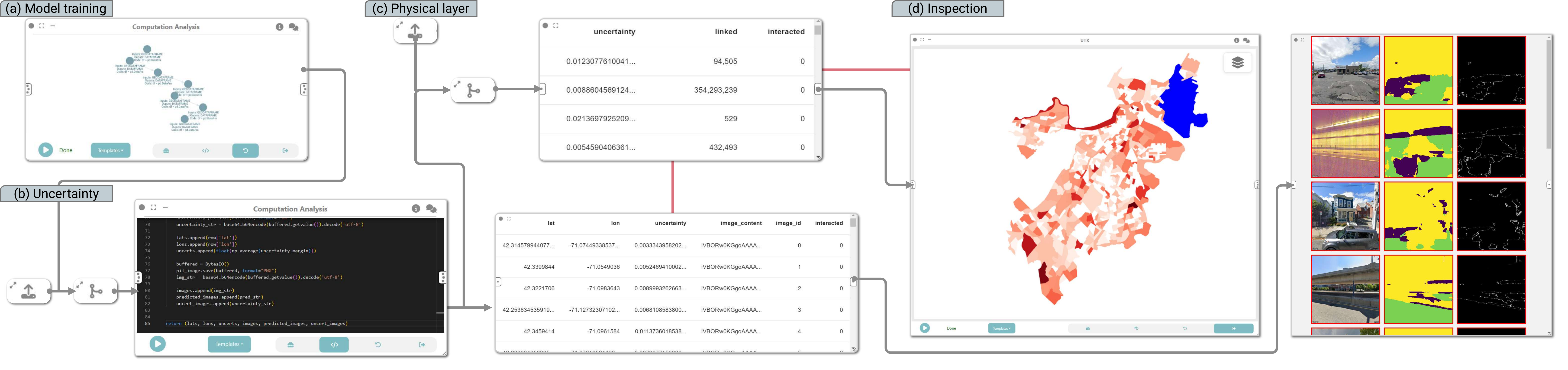}
\caption{Using \system to facilitate expert-in-the-loop inspection of a computer vision model. (a) The user begins by training the model. Provenance information is stored, allowing them to revert to previous versions of the model or explore different training parameters. \highlight{(b) New nodes are created to load unseen image data and compute the uncertainty of predictions. (c) A physical layer describing neighborhoods in Boston is loaded.} (d) An interactive visualization is created, enabling experts to analyze prediction uncertainty across neighborhoods in Boston.}
\vspace{-0.5cm}
\label{fig:active}
\end{figure*}

\subsection{Expert-in-the-loop urban accessibility analyses}

While urban livability and quality of life are highly dependent on well-designed public spaces, for a large group of urban dwellers, particularly those with mobility or vision impairments, these spaces remain out of reach. 
Thus, it is crucial to quantify the degree to which essential destinations are reachable by people with different levels of mobility~\cite{mason_universal_2017}.
In this context, the existence, quality, and surface material of sidewalks are major determinants of destination accessibility, specifically for elderly and wheelchair users~\cite{chippendale_neighborhood_2015}.
%
%
Despite their importance, few cities around the world maintain such spatial catalogues~\cite{deitz_squeaky_2021}. 
Recent advances in computer vision and the availability of street-level images have paved the way for low-cost, high-accuracy data collection on various built environment features, including sidewalk paving materials.

In this scenario, we illustrate how \system can facilitate the workflow of CitySurfaces~\cite{hosseini_citysurfaces_2022}, an active learning framework with expert-in-the-loop for the semantic segmentation of surface materials.
Active learning seeks to maximize accuracy while minimizing the number of required labeled data. By iteratively identifying and labeling the most informative or representative images, active learning reduces the number of necessary labeled instances to attain performance comparable to that achieved by labeling a large, randomly selected training dataset all at once~\cite{huang_active_2014}.
%
%
Since CitySurfaces tackles a challenging problem of in-the-wild texture segmentation with high within-class variation and between-class similarity, 
the training process should be carefully overseen by an expert to reduce systematic biases and identify patterns of failure and their spatial distribution~\cite{hosseini_citysurfaces_2022}. 


We use \system to automate the main part of the CitySurfaces workflow. An overview of the dataflow is presented in Figure~\ref{fig:active}.
E1 begins by importing their training procedure into a new \node{analysis \& modeling node} (a).
%
%
\system's provenance feature allows the expert to analyze several versions of their training procedure, also highlighted in (a).
After training, E1 creates an \node{analysis \& modeling node} with the procedures to calculate the difference between the two highest prediction probabilities in the Softmax layer (b).
Using \system's collaborative functionalities, a visualization researcher creates a \node{data node} to load the image data and joins it with Boston's neighborhoods (c).
A \node{UTK node} is then added to the dataflow with a spatial map showing the distribution of prediction uncertainties across Boston's neighborhoods (d).
%
%
%
%
%
%
%
These nodes are then used to identify potential shortcomings with the model, requiring new labeled data. The sorted mosaic of images also helps identify patterns of failures where the model had the most difficulty classifying.
This signals the need to sample more images with similar light, shadow, and built environment conditions.
%
%
%
Given that dense labeling of images is an expensive endeavor, \system facilitates a more targeted approach by identifying specific conditions that can guide the labeling process.

Importantly, the creation of this \system dataflow allows the expert to easily iterate over different stages, making use of interactive visualizations that link map and image gallery. 
This case illustrates how \system can significantly streamline the expert-in-the-loop process by creating a custom dataflow that integrates the sampling strategy into the main training pipeline, automatically computing the uncertainty metric. Furthermore, \system enables the visualization of  uncertainty maps to guide new sampling at each stage.
Tasks that were done in isolation in the original paper~\cite{hosseini_citysurfaces_2022} are now modularized into an easy-to-use and understandable dataflow.

\begin{figure*}[t!]
\centering
\includegraphics[width=1\linewidth]{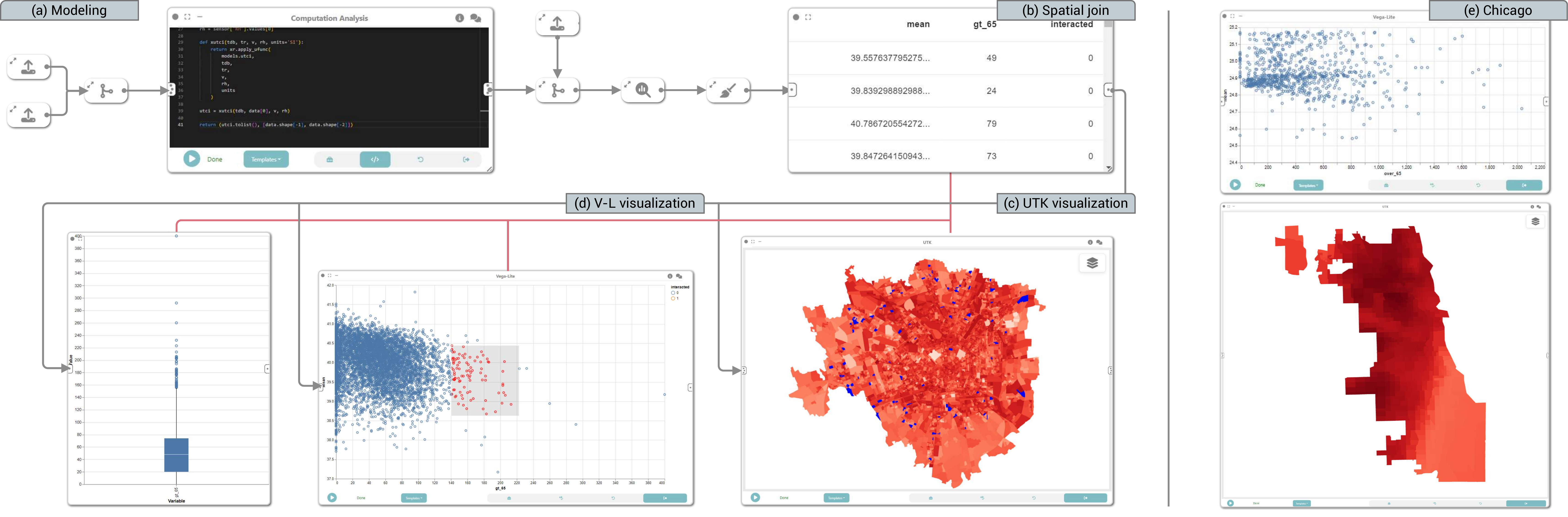}
\vspace{-0.5cm}
\caption{Using \system to create visualizations leveraging multiple datasets. (a) The user loads weather data and computes the UTCI, (b) followed by a spatial join with a physical layer describing neighborhoods in Milan. (c,d) The user then creates linked visualizations that highlight neighborhoods with high UTCI and a large population of older adults. (e) By changing two nodes, the user can create a similar visualization using data from Chicago.}
\vspace{-0.5cm}
\label{fig:heterogeneous}
\end{figure*}

\subsection{What-if scenario planning}

Urban development projects, particularly those involving the construction of high-rise buildings, pose significant challenges to the ecological balance and social fabric of cities. 
In Boston, the introduction of such developments has sparked serious debates~\cite{murphy_charles_2023}. At the heart of the controversy is the potential shadow cast by the new development on the Emerald Necklace, a cherished chain of parks in Boston. 
%
This situation has catalyzed community responses, leading to a petition urging the City of Boston to enforce and possibly extend its shadow protection policies~\cite{wolf_protect_2023}. Advocates for the parks argue that the new development threatens the vitality and accessibility of these valued green spaces. 
%
%
The profound ecological and societal ramifications of shadow accumulation on public green spaces demand a sophisticated, user-centric approach to urban planning and environmental stewardship.  

We use \system to create a dataflow that computes the shadow impact of the proposed buildings. Figure~\ref{fig:teaser} presents an overview of the dataflow.
%
With this dataflow, we can compute the shadow impact of the proposed buildings at different times and seasons to support evidence-based environmental impact analysis in Boston. 
%
%
The dataflow begins with a visualization researcher creating an \node{OSM node} that loads OpenStreetMap data from the region of interest (a).
A \node{sunlight access node} connected to the previous node computes the sunlight access (b), which is then visualized in a \node{UTK node} (c).
%
%

To support what-if scenarios, they add a custom \node{data transformation} node that receives the OSM data and changes the height of a selected building, generating an alternate scenario (d).
The shadow difference is shown in (e).
Such an application is made available to E1 in the visualization mode (f), along with GUI elements created through annotations (b,d). With the elements, they can interactively change the height of buildings using the GUI, and visualize the impact on the public space.
Importantly, all of the steps in this pipeline can be transparently accessed by stakeholders who wish to investigate the low-level functionalities.
E1 notes that the proposed buildings will add a considerable amount of shadow to the park.
Given \system's flexibility, E1 can quickly change the period and duration of shadow accumulation to gain a more comprehensive view of the proposed buildings' shadow impact.

By empowering researchers and stakeholders to conduct detailed before-and-after analyses of shadow impact assessments, \system takes a proactive approach to sustainable urban development by ensuring that the development's trajectory honors its commitment to ecological integrity and community well-being. The visualization mode can be made available to community residents, increasing visibility and engagement on this topic.

\subsection{Visual analytics of heterogeneous data}
\label{sec:heterogeneous}

As global temperatures continue to rise, urban areas are becoming increasingly susceptible to severe heat incidents, making sustainable active transportation modes less attractive~\cite{kim_effect_2022}. Outdoor thermal comfort and microclimate conditions in the urban environment are prime factors influencing the use of public spaces and can significantly impact the willingness to walk and bike~\cite{nikolopoulou_thermal_2001}.
In this scenario, we highlight how \system can be used to create dataflows for micro-scale environmental and human-scale analysis.
The objective is to demonstrate how urban planners and decision-makers can effectively use the framework to incorporate multiple datasets, including high-resolution microclimate variables, to assess heat stress levels, particularly for vulnerable populations.
Figure~\ref{fig:heterogeneous} provides an overview of the dataflow.

E2 first loads high-resolution mean radiant temperature data, stored as a TIFF file, using a \node{grid layer node}. 
Then, using a \node{file node}, E2 loads air temperature, relative humidity, and wind speed data from an ERA5 hourly meteorological dataset.
E2 then creates a custom \node{analysis \& modeling node} (a) that takes the previously loaded data as input and computes the Universal Thermal Climate Index (UTCI)~\cite{fiala_utci-fiala_2012}, a human biometeorology parameter used to assess human well-being in outdoor environments.

%
To study the impact of UTCI on vulnerable populations, particularly older adults, they create a new flow that loads sociodemographic data for adults over 65 at the neighborhood level and spatially join the UTCI data in raster format with the sociodemographic data (b). 
The results are visualized by adding a UTK map (c).
The visualization researcher then creates a linked scatterplot using a \node{Vega-Lite node} (d). The map and scatterplot are linked through an \node{interaction node}, allowing for the analysis of outliers of concern, i.e., regions that have large populations of older adults and high UTCI.

%
To highlight \system's flexibility in being adapted to other scenarios, we presented the dataflow to another urban expert (E3), who was responsible for porting the dataflow to Chicago (e). E3 only needed to change two data loading nodes to generate similar visualizations.
Further analyses could use a similar dataflow to identify individual routes that may expose vulnerable populations to higher temperatures.

\subsection{Experts' feedback}
After the creation of the dataflows, we met with the experts once more to get their perspectives regarding \system. We conducted semi-structured interviews, asking them about \system's strengths and limitations.
The experts responded positively to the ability to have a more transparent view of the dataflow steps. Compared to their usual workflow, E1 mentioned that the ability to interactively create a dataflow through a diagram model aligns with how they typically conceptualize their pipeline, describing it as a positive ``\emph{alternative to having a series of Python scripts lying around.}''

Regarding comparisons with existing approaches, E2 mentioned that ``\emph{the diagram interface resembled ArcGIS Pro's model builder}'', but added that ``\emph{\system offers more visualization options, and Python code integration is much easier.}''
E2 praised the diagram interface, saying that "\emph{we are used to working with visual programming, specifically those of us working extensively with Rhino's Grasshopper or ArcGIS Pro's model builder, which makes working with \system a breeze. Also, it does not tie the user to one specific software like Rhino and is much faster.}"
E2 also provided positive feedback about the ability to edit Python code, noting that, unlike off-the-shelf tools, ``\emph{you can optimize operations if you want to.}''
E3 was particularly impressed by the collaboration capabilities, stating that \system's ``\emph{main strength is the collaboration.}''

Regarding limitations, E2 mentioned that it would be important to have the ability to ``\emph{drag and drop CSV files from a folder onto the canvas, instead of creating a node to load files.}''
\highlight{Related to this point, E2 also mentioned that purely relying on grammars for the creation of visualizations might be cumbersome for some, as it requires experts to ``\emph{learn yet another software stack.}''}
The same expert was hesitant regarding adoption, mentioning that ``\emph{any framework needs a strong support system, with extensive documentation, examples, and an active user community.}''

\subsection{Reflection on design goals}
We now reflect on the initial design goals that guided our efforts.
Regarding DG1 (collaboration), \system supports asynchronous collaboration and includes several features to facilitate it. We believe that one of the most important collaborative features of the framework is the ability to link various facets of a node through annotations. This approach allowed for a much easier exploration of the parameter space in the second usage scenario.
Regarding DG2 (flexibility), \system supports the creation and connection of computing modules, as well as storage and retrieval of previously created ones. This was positively received by urban experts, particularly because they could easily visualize how the different components of their code connected with each other. \system's dataflow design enables it to address a wide range of urban-specific problems. As tool builders, we believe that a dataflow approach can lead to more modular and interoperable visual analytics systems.
Concerning DG3 (reproducibility and provenance), \system records changes made to a node in a provenance database. This database serves not only as a historical repository but also as a resource to visualize the evolution of a node over time. Users can leverage these visualizations to navigate through the different states of a node, enabling them to roll back to specific versions as needed. 


\hide{
Climate change and its associated increase in heat-related hazards pose a pressing threat to urban residents' health and well-being~\cite{?}. As the global temperature continues to rise, urban areas are increasingly susceptible to severe heat incidents, making more sustainable active transportation modes less attractive~\cite{kim2022effect}. Outdoor thermal comfort (OTC) and microclimate conditions of the urban environment are prime factors in the use of public spaces and hence can significantly impact the willingness to walk and bike~\cite{nikolopoulou2001thermal}.

Moving beyond traditional areal units like grids or census tracts, which do not effectively represent the micro-scale outdoor environment and human scales analysis, we incorporate a more precise measure of thermal conditions and adopt a sidewalk network-based analysis, offering a more realistic view of outdoor impacts during extreme heat events~\cite{colaninno2024urban}. \maryam{Should be removed Since we are not aggregating on sidewalk networks. }
The objective is to provide urban planners and decision-makers with a user-friendly and effective tool to simulate very-high-resolution (VHR) microclimate variables to inform the heat stress levels along the pedestrian network.
As a proof of concept, in this usage scenario, a high-resolution Mean Radiant Temperature (\Tmrt) map is used to generate the Universal Thermal Climate Index (UTCI)~\cite{fiala2012utci}. \Tmrt is an integral aspect of the Universal Thermal Climate Index (UTCI), which calculates the equivalent temperature felt by the human body, accounting for the combined effects of \Tmrt, air temperature (Ta), wind speed (Ws), and relative humidity (RH).

Typically, such analysis involves transitioning between various software and tools. The workflow often necessitates adequate selection and utilization of specialized software applications, as well as orchestration of the varied systems and their data requirement, to ensure consistency and operational efficiency, making it challenging to scale. 

Using \system, we streamline the computation process in a user-centric and interactive manner, promoting efficiency and accuracy in data processing and analysis. The interface empowers users to easily navigate, manipulate, and interpret a diverse array of computational tasks.

For this study, we estimate the UTCI across Milan, Italy, which covers an area of 181.67 km$^2$ and has a population of approximately 1.37 million. To estimate city-scale UTCI, we combine a Very High Resolution (VHI) \Tmrt map, generated leveraging the SOLWEIG (Solar and Longwave Environmental Irradiance Geometry) model~\cite{Lindberg2008}, with hourly Ta, Ws, and RH data provided by the ERA5 high-resolution meteorological dataset, produced by the Copernicus Climate Change Service (C3S) (see Figure~\ref{fig:street}). The resulting UTCI raster map is simulated for a typical hot summer day. 

The resulting map reveals areas in the city with higher intensity of thermal stress. To visualize the spatial distribution of thermal risk to vulnerable populations, we use \system to aggregate UTCI at the tract level and do a spatial join with the tract level demographic data. Next, we identify tracts with the highest populations of elderly (over 65) and children (under 5). We can then drill down to the street level and identify routes that can expose the vulnerable population to higher temperatures. 

%



}



\hide{
While urban liveability and quality of life are highly dependent on well-designed public spaces, for a large group of urban dwellers, particularly those with mobility or vision impairments, these spaces remain out of reach. Universal access, defined by the World Bank as ``the ability for people to reach the destinations necessary to lead productive and fulfilling lives" ~\cite{mason2018universal}, attempts to quantify the degree to which essential destinations are reachable by people with different level of mobility~\cite{}. The existence and quality of sidewalks are major determinants of destination accessibility. In this context, the surface material of sidewalks is of great importance, specifically for elderly and wheelchair users~\cite{jain2018recognizing}.

Despite its importance, few cities around the world maintain such spatial catalogues~\cite{}. Collecting data on the location and types of sidewalk surface materials using conventional, in-field methods is cost-prohibitive and not scalable~\cite{}. Recent advances in computer vision and the availability of street-level images paved the way for large-scale, low-cost, high-accuracy data collection on various built environment features, including sidewalk paving materials. 

In this use case, we illustrate how \system can be used to automate the workflow of CitySurfaces, an active learning framework with experts in the loop of training for the semantic segmentation of surface materials~\cite{hosseini2022citysurfaces}. 
Active learning seeks to maximize accuracy while minimizing the number of required labeled data. The underlying assumption is that a model can achieve superior performance with fewer labeled instances if it selects its training data~\cite{settles2009active}.
By iteratively identifying and labeling the most informative or representative images, active learning reduces the number of necessary labeled instances to attain performance comparable to that achieved by annotating a large, randomly selected training dataset in a single batch~\cite{huang2014active,bloodgood2014method}. 

CitySurfaces workflow has three major blocks:
(b) creating initial training labels; (b) training a material segmentation model and; (c) extending the model to segment three additional material classes from NYC standard materials. 
We use \system to automate the process in block B, which is the main part of the framework where active learning starts. The multi-stage active learning process, each stage comprising ten epochs works as follows: The training starts with an initial training set of images and annotated labels and the model is trained for one stage. At the end of each stage, the best-performing epoch with the highest mean Intersection over Union (mIoU) on the validation dataset is identified. Then, based on certain performance metrics from that stage, a new sample of unlabeled data is selected, annotated, and will then be added to the training data and the model goes to the next stage, where it is trained on a large training set including the initial and the new sets. 

The main sampling strategy used in our analysis is a multi-class uncertainty sampling known as margin sampling (MS)~\cite{scheffer2001active}. It calculates the difference between the two highest prediction probabilities on SoftMax to produce uncertainty maps. The smallest margin in each map is then chosen as the image-level uncertainty. We will then visualize the uncertainty maps for the top 20\% of the top failures of the model on the validation set, which shows the areas where the model had the most difficulty classifying. \maryam{then what happens here in your workflow? I used those images to do a similarity search over unlabeled data + those identified by the experts as having certain patterns and then annotated those by making an inference on them using the model and manually modifying the misclassifications. Not sure what happens here}
Since each image is also geo-tagged, we can visualize the distribution of uncertainty maps across the city to identify potential areas from which we should select more samples.\maryam{to me, this does not sound like a robust argument since the validation set does not have a balanced geographical distribution. }

This use case illustrates how using~\system can significantly help streamline the active learning process, integrating the sampling strategy into the main training pipeline by automatically computing the uncertainty metrics, and visualizing the uncertainty maps and confusion matrix to guide new sampling at each stage. Tasks that were done in isolation in the original paper, where automating this process was mentioned as an important future step~\cite{hosseini2022citysurfaces}.

}

\hide{
\subsubsection{Shadow Impact Assessment}


Urban development projects, particularly those involving the construction of high-rise buildings, pose significant challenges to the ecological balance and social fabric of cities. In Boston, MA, the introduction of such developments has ignited serious debates~\cite{BostonSun2023}. At the heart of the controversy is the potential shadow cast by the new development on the Emerald Necklace, a cherished chain of parks in Boston. The Emerald Necklace Conservancy has advocated for a comprehensive citywide shadow protection policy. Their concerns highlight the project's potential to disrupt the delicate balance of natural light essential for vegetation health and park usability, particularly emphasizing the adverse effects on waterways, landscapes, and the safety of park pathways during winter months. This situation has catalyzed community response, culminating in a petition that calls for the City of Boston to enforce and possibly extend its shadow protection policies~\cite{EmeraldNecklacePetition2023}. Advocates for the parks argue that the new development, with its proposed height far exceeding current guidelines, threatens the vitality and accessibility of these cherished green spaces. The petition represents a collective push for policy adherence and thoughtful urban development that respects Boston's historical parks and the broader ecological and social well-being of the city.
The profound ecological and social ramifications of shadow accumulation on public green spaces demand a sophisticated, user-centric approach to urban planning and environmental stewardship.  

Users can visually compare baseline and projected shadow scenarios through an intuitive interface, facilitating a deeper understanding of potential impacts on Boston's green spaces. Using \system, we can compute the shadow impact of the proposed building over different times and seasons to support evidence-based environmental impact analysis in Boston. 

\maryam{steps in the analysis}

By empowering researchers and stakeholders with the ability to conduct detailed before-and-after analyses of shadow impact assessment, \system takes a proactive approach to sustainable urban development by ensuring that Boston's development trajectory honors its commitment to ecological integrity and community well-being. 
}

\hide{
\subsubsection{Environmental justice}


Since the 1960s, social movements have highlighted the unequal environmental harm suffered by marginalized communities and minorities. In this context, it becomes essential for society to fight for environmental justice, which demands equal protection from environmental hazards and inclusive decision-making processes. 
Heat stress, accentuated by climate change, emerges as a prevalent environmental challenge, disproportionately affecting disadvantaged urban populations. Thus, understanding the current and future climate conditions is crucial in identifying associated injustices. This involves integrating several data sources, such as satellite imagery, weather forecasts, and sociodemographic data. 
However, managing and interpreting this multifaceted data is complex, which demands a visualization tool that helps domain experts throughout the process. This kind of tool must be accessible and user-friendly for several reasons. Firstly, to facilitate collaboration among experts with varying backgrounds, a common occurrence even within the climate community. Additionally, as every member of the public should play a role in environmental justice decision-making, it is essential to make them publicly available for communities to access and user-friendly enough for them to understand and trust.

In this case study, experts aim to explore temperature data from numerical models, satellite imagery, and sociodemographic data within the Chicago Metropolitan Region. Numerical weather models typically generate multiple variables stored in NetCDF format, while satellite data is commonly in raster format, and sociodemographic information is often organized into CSV files. Given this diversity, users must extract the desired variables from each source and convert them into CSV files, ensuring compatibility with the system. These files should contain geographic coordinates and corresponding values at each time stamp. Additionally, users may include OpenStreetMap (OSM) data and GeoJSON files to delineate the region of interest for exploration, which the system then converts into a binary format.

After the data preparation phase, the subsequent step involves defining the grammar specification. Here, users specify visualization parameters, interface layout, and preferences for data presentation. Following this, the system is initialized. It loads the provided CSV files, performs spatial joins and aggregations, and constructs the interface according to the grammar specification. Once the interface is ready, users can interact with it, selecting specific fields for exploration on a map-based visualization. They can choose from options such as temperature forecasts from numerical models, satellite data, or particular sociodemographic features, with the interface displaying a color-coded map based on the selected query. Additionally, a slider allows users to update the map according to different time stamps.

The system also supports map interactions, which are defined through the grammar specification. In this case study, users have specified that when a region of the map is clicked, a line chart will appear, showing various data trends for that specific region over time. These specifications, including the data source, type of map interactions, and their corresponding actions, can be easily modified through the grammar to suit different analytical needs.
}

\section{Conclusions}
In this paper, we introduced \system, a web-based framework designed to enhance the creation and execution of urban visual analytics workflows.
By providing an intuitive, shared platform with a robust suite of urban-specific computing modules, \system simplifies the integration, analysis, and visualization of complex urban data, supporting a broad spectrum of urban analytics tasks. 
\system is built with inclusivity in mind, accommodating users across various levels of expertise. This flexibility ensures that both domain experts and visualization researchers can contribute effectively, iterating on design choices to produce solutions that align with their specific requirements and expectations.
%

To anchor collaborative efforts in a robust framework, \system records each step in the dataflow's evolution. This not only safeguards the contributions of all participants, but also empowers users to explore alternative development trajectories, revisit previous states, and share comprehensive or partial dataflow with ease, thus enhancing reproducibility and transparency.
\system offers a robust, scalable platform for ongoing collaboration and innovation in urban visual analytics. 

\paragraph{Limitations}
\highlight{
%
First, \system's collaborative features are restricted to asynchronously commenting nodes, reusing and extending node templates, and browsing previous versions and states of nodes.
Second, it inherits the limitations of Vega-Lite and UTK. For example, it does not support audio or video data, two important types of urban data. 
Third, \system tracks modifications made by the user during the construction of a dataflow, but it does not track user interactions within visualization nodes.
Fourth, our current mechanism for coordination between visualizations is data-driven, meaning it is not possible to synchronize visualization states.
Lastly, our current implementation uses in-memory data structures to store and process datasets and does not perform rendering optimization operations.}

\paragraph{Future work}
\highlight{In future work, we plan to support synchronous collaboration sessions.
Moreover, although our evaluation methodology finds precedent in similar frameworks~\cite{yu_visflow_2017,lyi_gosling_2022}, we intend to perform a more in-depth evaluation by engaging experts across multiple domains to better assess our system's usability and learnability.
With respect to provenance features, we also plan to extend Curio to support the provenance of interactions within visualization nodes.
Regarding the support of data types, we aim to explore approaches to integrate other complex data, such as audio and video, into the framework.
We also plan to revisit the use of dataflows as an approach for visualization education~\cite{silva_using_2011}, focusing on urban data and societal problems.}

\section*{Acknowledgments}
We would like to thank the reviewers for their constructive comments and feedback. This study was supported by the Discovery Partners Institute, the National Science Foundation (\#2320261, \#2330565, \#2411223), NASA (\#80NSSC22K1683), IDOT (TS-22-340), CNPq (316963/2021-6, 311425/2023-2), and FAPERJ (E-26/202.915/2019, E-26/211.134/2019).

\bibliographystyle{abbrv-doi-hyperref}

\bibliography{zotero}

\begin{thebibliography}{10}

\bibitem{akbaba_troubling_2023}
D.~Akbaba, D.~Lange, M.~Correll, A.~Lex, and M.~Meyer.
\newblock Troubling collaboration: {Matters} of care for visualization design study.
\newblock In {\em Proceedings of the {CHI} {Conference} on {Human} {Factors} in {Computing} {Systems}}, pp. 1--15. ACM, 2023. \href{https://doi.org/10.1145/3544548.3581168}
{doi: {{%
10\hspace{.1pt}\discretionary{.}{%
}{.}\hspace{.4pt}1145\discretionary{/}{%
}{/}3544548\hspace{.1pt}\discretionary{.}{%
}{.}\hspace{.4pt}3581168}}}


\bibitem{alam_survey_2022}
M.~M. Alam, L.~Torgo, and A.~Bifet.
\newblock A survey on spatio-temporal data analytics systems.
\newblock {\em ACM Computing Surveys}, 54(10s):1--38, 2022. \href{https://doi.org/10.1145/3507904}
{doi: {{%
10\hspace{.1pt}\discretionary{.}{%
}{.}\hspace{.4pt}1145\discretionary{/}{%
}{/}3507904}}}


\bibitem{alspaugh_futzing_2019}
S.~Alspaugh, N.~Zokaei, A.~Liu, C.~Jin, and M.~A. Hearst.
\newblock Futzing and moseying: {Interviews} with professional data analysts on exploration practices.
\newblock {\em IEEE Transactions on Visualization and Computer Graphics}, 25(1):22--31, 2019. \href{https://doi.org/10.1109/TVCG.2018.2865040}
{doi: {{%
10\hspace{.1pt}\discretionary{.}{%
}{.}\hspace{.4pt}1109\discretionary{/}{%
}{/}TVCG\hspace{.1pt}\discretionary{.}{%
}{.}\hspace{.4pt}2018\hspace{.1pt}\discretionary{.}{%
}{.}\hspace{.4pt}2865040}}}


\bibitem{ang_big_2016}
L.~Ang and K.~P. Seng.
\newblock Big sensor data applications in urban environments.
\newblock {\em Big Data Research}, 4:1--12, 2016. \href{https://doi.org/10.1016/j.bdr.2015.12.003}
{doi: {{%
10\hspace{.1pt}\discretionary{.}{%
}{.}\hspace{.4pt}1016\discretionary{/}{%
}{/}j\hspace{.1pt}\discretionary{.}{%
}{.}\hspace{.4pt}bdr\hspace{.1pt}\discretionary{.}{%
}{.}\hspace{.4pt}2015\hspace{.1pt}\discretionary{.}{%
}{.}\hspace{.4pt}12\hspace{.1pt}\discretionary{.}{%
}{.}\hspace{.4pt}003}}}


\bibitem{barbosa_structured_2014}
L.~Barbosa, K.~Pham, C.~Silva, M.~R. Vieira, and J.~Freire.
\newblock Structured open urban data: {Understanding} the landscape.
\newblock {\em Big Data}, 2(3):144--154, 2014. \href{https://doi.org/10.1089/big.2014.0020}
{doi: {{%
10\hspace{.1pt}\discretionary{.}{%
}{.}\hspace{.4pt}1089\discretionary{/}{%
}{/}big\hspace{.1pt}\discretionary{.}{%
}{.}\hspace{.4pt}2014\hspace{.1pt}\discretionary{.}{%
}{.}\hspace{.4pt}0020}}}


\bibitem{batch_interactive_2018}
A.~Batch and N.~Elmqvist.
\newblock The interactive visualization gap in initial exploratory data analysis.
\newblock {\em IEEE Transactions on Visualization and Computer Graphics}, 24(1):278--287, 2018. \href{https://doi.org/10.1109/TVCG.2017.2743990}
{doi: {{%
10\hspace{.1pt}\discretionary{.}{%
}{.}\hspace{.4pt}1109\discretionary{/}{%
}{/}TVCG\hspace{.1pt}\discretionary{.}{%
}{.}\hspace{.4pt}2017\hspace{.1pt}\discretionary{.}{%
}{.}\hspace{.4pt}2743990}}}


\bibitem{bavoil_vistrails_2005}
L.~Bavoil, S.~Callahan, P.~Crossno, J.~Freire, C.~Scheidegger, C.~Silva, and H.~Vo.
\newblock {VisTrails}: {Enabling} interactive multiple-view visualizations.
\newblock In {\em {IEEE} {Visualization}}, pp. 135--142. IEEE, 2005. \href{https://doi.org/10.1109/VISUAL.2005.1532788}
{doi: {{%
10\hspace{.1pt}\discretionary{.}{%
}{.}\hspace{.4pt}1109\discretionary{/}{%
}{/}VISUAL\hspace{.1pt}\discretionary{.}{%
}{.}\hspace{.4pt}2005\hspace{.1pt}\discretionary{.}{%
}{.}\hspace{.4pt}1532788}}}


\bibitem{biljecki_street_2021}
F.~Biljecki and K.~Ito.
\newblock Street view imagery in urban analytics and {GIS}: {A} review.
\newblock {\em Landscape and Urban Planning}, 215:104217, 2021. \href{https://doi.org/10.1016/j.landurbplan.2021.104217}
{doi: {{%
10\hspace{.1pt}\discretionary{.}{%
}{.}\hspace{.4pt}1016\discretionary{/}{%
}{/}j\hspace{.1pt}\discretionary{.}{%
}{.}\hspace{.4pt}landurbplan\hspace{.1pt}\discretionary{.}{%
}{.}\hspace{.4pt}2021\hspace{.1pt}\discretionary{.}{%
}{.}\hspace{.4pt}104217}}}


\bibitem{bin_multi-source_2020}
J.~Bin, B.~Gardiner, E.~Li, and Z.~Liu.
\newblock Multi-source urban data fusion for property value assessment: {A} case study in {Philadelphia}.
\newblock {\em Neurocomputing}, 404:70--83, 2020. \href{https://doi.org/10.1016/j.neucom.2020.05.013}
{doi: {{%
10\hspace{.1pt}\discretionary{.}{%
}{.}\hspace{.4pt}1016\discretionary{/}{%
}{/}j\hspace{.1pt}\discretionary{.}{%
}{.}\hspace{.4pt}neucom\hspace{.1pt}\discretionary{.}{%
}{.}\hspace{.4pt}2020\hspace{.1pt}\discretionary{.}{%
}{.}\hspace{.4pt}05\hspace{.1pt}\discretionary{.}{%
}{.}\hspace{.4pt}013}}}


\bibitem{chippendale_neighborhood_2015}
T.~Chippendale and M.~Boltz.
\newblock The neighborhood environment: {Perceived} fall risk, resources, and strategies for fall prevention.
\newblock {\em The Gerontologist}, 55(4):575--583, 2015. \href{https://doi.org/10.1093/geront/gnu019}
{doi: {{%
10\hspace{.1pt}\discretionary{.}{%
}{.}\hspace{.4pt}1093\discretionary{/}{%
}{/}geront\discretionary{/}{%
}{/}gnu019}}}


\bibitem{degbelo_fair_2021}
A.~Degbelo.
\newblock {FAIR} geovisualizations: {Definitions}, challenges, and the road ahead.
\newblock {\em International Journal of Geographical Information Science}, 36(6):1059--1099, 2021. \href{https://doi.org/10.1080/13658816.2021.1983579}
{doi: {{%
10\hspace{.1pt}\discretionary{.}{%
}{.}\hspace{.4pt}1080\discretionary{/}{%
}{/}13658816\hspace{.1pt}\discretionary{.}{%
}{.}\hspace{.4pt}2021\hspace{.1pt}\discretionary{.}{%
}{.}\hspace{.4pt}1983579}}}


\bibitem{deitz_squeaky_2021}
S.~Deitz, A.~Lobben, and A.~Alferez.
\newblock Squeaky wheels: {Missing} data, disability, and power in the smart city.
\newblock {\em Big Data \& Society}, 8(2):1--16, 2021. \href{https://doi.org/10.1177/20539517211047735}
{doi: {{%
10\hspace{.1pt}\discretionary{.}{%
}{.}\hspace{.4pt}1177\discretionary{/}{%
}{/}20539517211047735}}}


\bibitem{deng_airvis_2020}
Z.~Deng, D.~Weng, J.~Chen, R.~Liu, Z.~Wang, J.~Bao, Y.~Zheng, and Y.~Wu.
\newblock {AirVis}: {Visual} analytics of air pollution propagation.
\newblock {\em IEEE Transactions on Visualization and Computer Graphics}, 26(1):800--810, 2020. \href{https://doi.org/10.1109/TVCG.2019.2934670}
{doi: {{%
10\hspace{.1pt}\discretionary{.}{%
}{.}\hspace{.4pt}1109\discretionary{/}{%
}{/}TVCG\hspace{.1pt}\discretionary{.}{%
}{.}\hspace{.4pt}2019\hspace{.1pt}\discretionary{.}{%
}{.}\hspace{.4pt}2934670}}}


\bibitem{deng_survey_2023}
Z.~Deng, D.~Weng, S.~Liu, Y.~Tian, M.~Xu, and Y.~Wu.
\newblock A survey of urban visual analytics: {Advances} and future directions.
\newblock {\em Computational Visual Media}, 9:3--39, 2023. \href{https://doi.org/10.1007/s41095-022-0275-7}
{doi: {{%
10\hspace{.1pt}\discretionary{.}{%
}{.}\hspace{.4pt}1007\discretionary{/}{%
}{/}s41095\discretionary{%
}{-}{-}022\discretionary{%
}{-}{-}0275\discretionary{%
}{-}{-}7}}}


\bibitem{doraiswamy_interactive_2018}
H.~Doraiswamy, E.~Tzirita~Zacharatou, F.~Miranda, M.~Lage, A.~Ailamaki, C.~T. Silva, and J.~Freire.
\newblock Interactive visual exploration of spatio-temporal urban data sets using {Urbane}.
\newblock In {\em International {Conference} on {Management} of {Data}}, pp. 1693--1696. ACM, 2018. \href{https://doi.org/10.1145/3183713.3193559}
{doi: {{%
10\hspace{.1pt}\discretionary{.}{%
}{.}\hspace{.4pt}1145\discretionary{/}{%
}{/}3183713\hspace{.1pt}\discretionary{.}{%
}{.}\hspace{.4pt}3193559}}}


\bibitem{feng_survey_2022}
Z.~Feng, H.~Qu, S.-H. Yang, Y.~Ding, and J.~Song.
\newblock A survey of visual analytics in urban area.
\newblock {\em Expert Systems}, 39(9):e13065, 2022. \href{https://doi.org/10.1111/exsy.13065}
{doi: {{%
10\hspace{.1pt}\discretionary{.}{%
}{.}\hspace{.4pt}1111\discretionary{/}{%
}{/}exsy\hspace{.1pt}\discretionary{.}{%
}{.}\hspace{.4pt}13065}}}


\bibitem{ferreira_assessing_2024}
L.~Ferreira, G.~Moreira, M.~Hosseini, M.~Lage, N.~Ferreira, and F.~Miranda.
\newblock Assessing the landscape of toolkits, frameworks, and authoring tools for urban visual analytics systems.
\newblock {\em Computers \& Graphics}, 123:104013, 2024. \href{https://doi.org/10.1016/j.cag.2024.104013}
{doi: {{%
10\hspace{.1pt}\discretionary{.}{%
}{.}\hspace{.4pt}1016\discretionary{/}{%
}{/}j\hspace{.1pt}\discretionary{.}{%
}{.}\hspace{.4pt}cag\hspace{.1pt}\discretionary{.}{%
}{.}\hspace{.4pt}2024\hspace{.1pt}\discretionary{.}{%
}{.}\hspace{.4pt}104013}}}


\bibitem{ferreira_urbane_2015}
N.~Ferreira, M.~Lage, H.~Doraiswamy, H.~Vo, L.~Wilson, H.~Werner, M.~Park, and C.~Silva.
\newblock Urbane: {A} {3D} framework to support data driven decision making in urban development.
\newblock In {\em Conference on {Visual} {Analytics} {Science} and {Technology}}, pp. 97--104. IEEE, 2015. \href{https://doi.org/10.1109/VAST.2015.7347636}
{doi: {{%
10\hspace{.1pt}\discretionary{.}{%
}{.}\hspace{.4pt}1109\discretionary{/}{%
}{/}VAST\hspace{.1pt}\discretionary{.}{%
}{.}\hspace{.4pt}2015\hspace{.1pt}\discretionary{.}{%
}{.}\hspace{.4pt}7347636}}}


\bibitem{ferreira_visual_2013}
N.~Ferreira, J.~Poco, H.~T. Vo, J.~Freire, and C.~T. Silva.
\newblock Visual exploration of big spatio-temporal urban data: {A} study of {New York City} taxi trips.
\newblock {\em IEEE Transactions on Visualization and Computer Graphics}, 19(12):2149--2158, 2013. \href{https://doi.org/10.1109/TVCG.2013.226}
{doi: {{%
10\hspace{.1pt}\discretionary{.}{%
}{.}\hspace{.4pt}1109\discretionary{/}{%
}{/}TVCG\hspace{.1pt}\discretionary{.}{%
}{.}\hspace{.4pt}2013\hspace{.1pt}\discretionary{.}{%
}{.}\hspace{.4pt}226}}}


\bibitem{fiala_utci-fiala_2012}
D.~Fiala, G.~Havenith, P.~Bröde, B.~Kampmann, and G.~Jendritzky.
\newblock {UTCI}-{Fiala} multi-node model of human heat transfer and temperature regulation.
\newblock {\em International Journal of Biometeorology}, 56:429--441, 2012. \href{https://doi.org/10.1007/s00484-011-0424-7}
{doi: {{%
10\hspace{.1pt}\discretionary{.}{%
}{.}\hspace{.4pt}1007\discretionary{/}{%
}{/}s00484\discretionary{%
}{-}{-}011\discretionary{%
}{-}{-}0424\discretionary{%
}{-}{-}7}}}


\bibitem{groth_w3c_2013}
P.~Groth and L.~Moreau.
\newblock {W3C} {PROV} - {An} overview of the {PROV} family of documents.
\newblock \url{https://www.w3.org/TR/prov-overview/}, 2013.
\newblock Accessed on: Jun 2024.

\bibitem{grus_i_2018}
J.~Grus.
\newblock I don't like notebooks.
\newblock \url{https://conferences.oreilly.com/jupyter/jup-ny/public/schedule/detail/68282.html}, 2018.
\newblock Accessed on: Jun 2024.

\bibitem{haeberli_conman_1988}
P.~E. Haeberli.
\newblock {ConMan}: {A} visual programming language for interactive graphics.
\newblock In {\em Annual {Conference} on {Computer} {Graphics} and {Interactive} {Techniques}}, pp. 103--111. ACM, 1988. \href{https://doi.org/10.1145/54852.378494}
{doi: {{%
10\hspace{.1pt}\discretionary{.}{%
}{.}\hspace{.4pt}1145\discretionary{/}{%
}{/}54852\hspace{.1pt}\discretionary{.}{%
}{.}\hspace{.4pt}378494}}}


\bibitem{hanwell_visualization_2015}
M.~D. Hanwell, K.~M. Martin, A.~Chaudhary, and L.~S. Avila.
\newblock The {Visualization} {Toolkit} ({VTK}): {Rewriting} the rendering code for modern graphics cards.
\newblock {\em SoftwareX}, 1:9--12, 2015. \href{https://doi.org/10.1016/j.softx.2015.04.001}
{doi: {{%
10\hspace{.1pt}\discretionary{.}{%
}{.}\hspace{.4pt}1016\discretionary{/}{%
}{/}j\hspace{.1pt}\discretionary{.}{%
}{.}\hspace{.4pt}softx\hspace{.1pt}\discretionary{.}{%
}{.}\hspace{.4pt}2015\hspace{.1pt}\discretionary{.}{%
}{.}\hspace{.4pt}04\hspace{.1pt}\discretionary{.}{%
}{.}\hspace{.4pt}001}}}


\bibitem{herschel1_survey_2017}
M.~Herschel, R.~Diestelkämper, and H.~Lahmar.
\newblock A survey on provenance: {What} for? {What} form? {What} from?
\newblock {\em The VLDB Journal}, 26:881--906, 2017. \href{https://doi.org/10.1007/s00778-017-0486-1}
{doi: {{%
10\hspace{.1pt}\discretionary{.}{%
}{.}\hspace{.4pt}1007\discretionary{/}{%
}{/}s00778\discretionary{%
}{-}{-}017\discretionary{%
}{-}{-}0486\discretionary{%
}{-}{-}1}}}


\bibitem{hosseini_citysurfaces_2022}
M.~Hosseini, F.~Miranda, J.~Lin, and C.~Silva.
\newblock {CitySurfaces}: {City}-scale semantic segmentation of sidewalk materials.
\newblock {\em Sustainable Cities and Society}, 79:103630, 2022. \href{https://doi.org/10.1016/j.scs.2021.103630}
{doi: {{%
10\hspace{.1pt}\discretionary{.}{%
}{.}\hspace{.4pt}1016\discretionary{/}{%
}{/}j\hspace{.1pt}\discretionary{.}{%
}{.}\hspace{.4pt}scs\hspace{.1pt}\discretionary{.}{%
}{.}\hspace{.4pt}2021\hspace{.1pt}\discretionary{.}{%
}{.}\hspace{.4pt}103630}}}


\bibitem{huang_active_2014}
S.-J. Huang, R.~Jin, and Z.-H. Zhou.
\newblock Active learning by querying informative and representative examples.
\newblock {\em IEEE Transactions on Pattern Analysis and Machine Intelligence}, 36(10):1936--1949, 2014. \href{https://doi.org/10.1109/TPAMI.2014.2307881}
{doi: {{%
10\hspace{.1pt}\discretionary{.}{%
}{.}\hspace{.4pt}1109\discretionary{/}{%
}{/}TPAMI\hspace{.1pt}\discretionary{.}{%
}{.}\hspace{.4pt}2014\hspace{.1pt}\discretionary{.}{%
}{.}\hspace{.4pt}2307881}}}


\bibitem{ikeda_logical_2013}
R.~Ikeda, A.~Das~Sarma, and J.~Widom.
\newblock Logical provenance in data-oriented workflows?
\newblock In {\em International {Conference} on {Data} {Engineering}}, pp. 877--888. IEEE, 2013. \href{https://doi.org/10.1109/ICDE.2013.6544882}
{doi: {{%
10\hspace{.1pt}\discretionary{.}{%
}{.}\hspace{.4pt}1109\discretionary{/}{%
}{/}ICDE\hspace{.1pt}\discretionary{.}{%
}{.}\hspace{.4pt}2013\hspace{.1pt}\discretionary{.}{%
}{.}\hspace{.4pt}6544882}}}


\bibitem{isenberg_collaborative_2011}
P.~Isenberg, N.~Elmqvist, J.~Scholtz, D.~Cernea, K.-L. Ma, and H.~Hagen.
\newblock Collaborative visualization: {Definition}, challenges, and research agenda.
\newblock {\em Information Visualization}, 10(4):310--326, 2011. \href{https://doi.org/10.1177/1473871611412817}
{doi: {{%
10\hspace{.1pt}\discretionary{.}{%
}{.}\hspace{.4pt}1177\discretionary{/}{%
}{/}1473871611412817}}}


\bibitem{javed_explates_2013}
W.~Javed and N.~Elmqvist.
\newblock {ExPlates}: {Spatializing} interactive analysis to scaffold visual exploration.
\newblock {\em Computer Graphics Forum}, 32(3pt4):441--450, 2013. \href{https://doi.org/10.1111/cgf.12131}
{doi: {{%
10\hspace{.1pt}\discretionary{.}{%
}{.}\hspace{.4pt}1111\discretionary{/}{%
}{/}cgf\hspace{.1pt}\discretionary{.}{%
}{.}\hspace{.4pt}12131}}}


\bibitem{janicke_participatory_2020}
S.~Jänicke, P.~Kaur, P.~Kuzmicki, and J.~Schmidt.
\newblock Participatory visualization design as an approach to minimize the gap between research and application.
\newblock In {\em {VisGap} - {The} {Gap} {Between} {Visualization} {Research} and {Visualization} {Software}}. EG, 2020. \href{https://doi.org/10.2312/visgap.20201108}
{doi: {{%
10\hspace{.1pt}\discretionary{.}{%
}{.}\hspace{.4pt}2312\discretionary{/}{%
}{/}visgap\hspace{.1pt}\discretionary{.}{%
}{.}\hspace{.4pt}20201108}}}


\bibitem{kandt_smart_2021}
J.~Kandt and M.~Batty.
\newblock Smart cities, big data and urban policy: {Towards} urban analytics for the long run.
\newblock {\em Cities}, 109:102992, 2021. \href{https://doi.org/10.1016/j.cities.2020.102992}
{doi: {{%
10\hspace{.1pt}\discretionary{.}{%
}{.}\hspace{.4pt}1016\discretionary{/}{%
}{/}j\hspace{.1pt}\discretionary{.}{%
}{.}\hspace{.4pt}cities\hspace{.1pt}\discretionary{.}{%
}{.}\hspace{.4pt}2020\hspace{.1pt}\discretionary{.}{%
}{.}\hspace{.4pt}102992}}}


\bibitem{kim_effect_2022}
Y.~Kim and R.~Brown.
\newblock Effect of meteorological conditions on leisure walking: {A} time series analysis and the application of outdoor thermal comfort indexes.
\newblock {\em International Journal of Biometeorology}, 66(6):1109--1123, 2022. \href{https://doi.org/10.1007/s00484-022-02262-w}
{doi: {{%
10\hspace{.1pt}\discretionary{.}{%
}{.}\hspace{.4pt}1007\discretionary{/}{%
}{/}s00484\discretionary{%
}{-}{-}022\discretionary{%
}{-}{-}02262\discretionary{%
}{-}{-}w}}}


\bibitem{knuth_literate_1984}
D.~E. Knuth.
\newblock Literate programming.
\newblock {\em The Computer Journal}, 27(2):97--111, 1984. \href{https://doi.org/10.1093/comjnl/27.2.97}
{doi: {{%
10\hspace{.1pt}\discretionary{.}{%
}{.}\hspace{.4pt}1093\discretionary{/}{%
}{/}comjnl\discretionary{/}{%
}{/}27\hspace{.1pt}\discretionary{.}{%
}{.}\hspace{.4pt}2\hspace{.1pt}\discretionary{.}{%
}{.}\hspace{.4pt}97}}}


\bibitem{kontokosta_urban_2021}
C.~E. Kontokosta.
\newblock Urban informatics in the science and practice of planning.
\newblock {\em Journal of Planning Education and Research}, 41(4):382--395, 2021. \href{https://doi.org/10.1177/0739456X18793716}
{doi: {{%
10\hspace{.1pt}\discretionary{.}{%
}{.}\hspace{.4pt}1177\discretionary{/}{%
}{/}0739456X18793716}}}


\bibitem{kunze_visualization_2012}
A.~Kunze, R.~Burkhard, S.~Gebhardt, and B.~Tuncer.
\newblock Visualization and decision support tools in urban planning.
\newblock In {\em Digital {Urban} {Modeling} and {Simulation}}, pp. 279--298. Springer, 2012. \href{https://doi.org/10.1007/978-3-642-29758-8_15}
{doi: {{%
10\hspace{.1pt}\discretionary{.}{%
}{.}\hspace{.4pt}1007\discretionary{/}{%
}{/}978\discretionary{%
}{-}{-}3\discretionary{%
}{-}{-}642\discretionary{%
}{-}{-}29758\discretionary{%
}{-}{-}8\_15}}}


\bibitem{lee_visual_2020}
C.~Lee, Y.~Kim, S.~Jin, D.~Kim, R.~Maciejewski, D.~Ebert, and S.~Ko.
\newblock A visual analytics system for exploring, monitoring, and forecasting road traffic congestion.
\newblock {\em IEEE Transactions on Visualization and Computer Graphics}, 26(11):3133--3146, 2020. \href{https://doi.org/10.1109/TVCG.2019.2922597}
{doi: {{%
10\hspace{.1pt}\discretionary{.}{%
}{.}\hspace{.4pt}1109\discretionary{/}{%
}{/}TVCG\hspace{.1pt}\discretionary{.}{%
}{.}\hspace{.4pt}2019\hspace{.1pt}\discretionary{.}{%
}{.}\hspace{.4pt}2922597}}}


\bibitem{loibl_effects_2021}
W.~Loibl, M.~Vuckovic, G.~Etminan, M.~Ratheiser, S.~Tschannett, and D.~Österreicher.
\newblock Effects of densification on urban microclimate—a case study for the city of {Vienna}.
\newblock {\em Atmosphere}, 12(4):1--23, 2021. \href{https://doi.org/10.3390/atmos12040511}
{doi: {{%
10\hspace{.1pt}\discretionary{.}{%
}{.}\hspace{.4pt}3390\discretionary{/}{%
}{/}atmos12040511}}}


\bibitem{luca_crime_2023}
M.~Luca, G.~M. Campedelli, S.~Centellegher, M.~Tizzoni, and B.~Lepri.
\newblock Crime, inequality and public health: {A} survey of emerging trends in urban data science.
\newblock {\em Frontiers in Big Data}, 6:1--20, 2023. \href{https://doi.org/10.3389/fdata.2023.1124526}
{doi: {{%
10\hspace{.1pt}\discretionary{.}{%
}{.}\hspace{.4pt}3389\discretionary{/}{%
}{/}fdata\hspace{.1pt}\discretionary{.}{%
}{.}\hspace{.4pt}2023\hspace{.1pt}\discretionary{.}{%
}{.}\hspace{.4pt}1124526}}}


\bibitem{lukasczyk_collaborative_2015}
J.~Lukasczyk, X.~Liang, W.~Luo, E.~D. Ragan, A.~Middel, N.~Bliss, D.~White, H.~Hagen, and R.~Maciejewski.
\newblock A collaborative web-based environmental data visualization and analysis framework.
\newblock In {\em Workshop on {Visualisation} in {Environmental} {Sciences}}. EG, 2015. \href{https://doi.org/10.2312/envirvis.20151087}
{doi: {{%
10\hspace{.1pt}\discretionary{.}{%
}{.}\hspace{.4pt}2312\discretionary{/}{%
}{/}envirvis\hspace{.1pt}\discretionary{.}{%
}{.}\hspace{.4pt}20151087}}}


\bibitem{lyi_gosling_2022}
S.~L'Yi, Q.~Wang, F.~Lekschas, and N.~Gehlenborg.
\newblock Gosling: {A} grammar-based toolkit for scalable and interactive genomics data visualization.
\newblock {\em IEEE Transactions on Visualization and Computer Graphics}, 28(1):140--150, 2022. \href{https://doi.org/10.1109/TVCG.2021.3114876}
{doi: {{%
10\hspace{.1pt}\discretionary{.}{%
}{.}\hspace{.4pt}1109\discretionary{/}{%
}{/}TVCG\hspace{.1pt}\discretionary{.}{%
}{.}\hspace{.4pt}2021\hspace{.1pt}\discretionary{.}{%
}{.}\hspace{.4pt}3114876}}}


\bibitem{lyu_if-city_2025}
Y.~Lyu, H.~Lu, M.~K. Lee, G.~Schmitt, and B.~Y. Lim.
\newblock {IF}-{City}: {Intelligible} fair city planning to measure, explain and mitigate inequality.
\newblock {\em IEEE Transactions on Visualization and Computer Graphics}, 30(7):3749--3766, 2024. \href{https://doi.org/10.1109/TVCG.2023.3239909}
{doi: {{%
10\hspace{.1pt}\discretionary{.}{%
}{.}\hspace{.4pt}1109\discretionary{/}{%
}{/}TVCG\hspace{.1pt}\discretionary{.}{%
}{.}\hspace{.4pt}2023\hspace{.1pt}\discretionary{.}{%
}{.}\hspace{.4pt}3239909}}}


\bibitem{mason_universal_2017}
J.~Mason, P.~Turner, and M.~Steriu.
\newblock Universal access in urban areas: {Why} universal access in urban areas matters for sustainable mobility.
\newblock Technical report, World Bank Group, Washington, D.C., 2017.

\bibitem{mei_design_2018}
H.~Mei, Y.~Ma, Y.~Wei, and W.~Chen.
\newblock The design space of construction tools for information visualization: {A} survey.
\newblock {\em Journal of Visual Languages \& Computing}, 44:120--132, 2018. \href{https://doi.org/10.1016/j.jvlc.2017.10.001}
{doi: {{%
10\hspace{.1pt}\discretionary{.}{%
}{.}\hspace{.4pt}1016\discretionary{/}{%
}{/}j\hspace{.1pt}\discretionary{.}{%
}{.}\hspace{.4pt}jvlc\hspace{.1pt}\discretionary{.}{%
}{.}\hspace{.4pt}2017\hspace{.1pt}\discretionary{.}{%
}{.}\hspace{.4pt}10\hspace{.1pt}\discretionary{.}{%
}{.}\hspace{.4pt}001}}}


\bibitem{miranda_shadow_2019}
F.~Miranda, H.~Doraiswamy, M.~Lage, L.~Wilson, M.~Hsieh, and C.~T. Silva.
\newblock {Shadow Accrual Maps}: {Efficient} accumulation of city-scale shadows over time.
\newblock {\em IEEE Transactions on Visualization and Computer Graphics}, 25(3):1559--1574, 2019. \href{https://doi.org/10.1109/TVCG.2018.2802945}
{doi: {{%
10\hspace{.1pt}\discretionary{.}{%
}{.}\hspace{.4pt}1109\discretionary{/}{%
}{/}TVCG\hspace{.1pt}\discretionary{.}{%
}{.}\hspace{.4pt}2018\hspace{.1pt}\discretionary{.}{%
}{.}\hspace{.4pt}2802945}}}


\bibitem{miranda_urban_2017}
F.~Miranda, H.~Doraiswamy, M.~Lage, K.~Zhao, B.~Gonçalves, L.~Wilson, M.~Hsieh, and C.~T. Silva.
\newblock {Urban Pulse}: {Capturing} the rhythm of cities.
\newblock {\em IEEE Transactions on Visualization and Computer Graphics}, 23(1):791--800, 2017. \href{https://doi.org/10.1109/TVCG.2016.2598585}
{doi: {{%
10\hspace{.1pt}\discretionary{.}{%
}{.}\hspace{.4pt}1109\discretionary{/}{%
}{/}TVCG\hspace{.1pt}\discretionary{.}{%
}{.}\hspace{.4pt}2016\hspace{.1pt}\discretionary{.}{%
}{.}\hspace{.4pt}2598585}}}


\bibitem{miranda_urban_2020}
F.~Miranda, M.~Hosseini, M.~Lage, H.~Doraiswamy, G.~Dove, and C.~T. Silva.
\newblock Urban {Mosaic}: {Visual} exploration of streetscapes using large-scale image data.
\newblock In {\em Proceedings of the {CHI} {Conference} on {Human} {Factors} in {Computing} {Systems}}, pp. 1--15. ACM, 2020. \href{https://doi.org/10.1145/3313831.3376399}
{doi: {{%
10\hspace{.1pt}\discretionary{.}{%
}{.}\hspace{.4pt}1145\discretionary{/}{%
}{/}3313831\hspace{.1pt}\discretionary{.}{%
}{.}\hspace{.4pt}3376399}}}


\bibitem{miranda_star_2024}
F.~Miranda, T.~Ortner, G.~Moreira, M.~Hosseini, M.~Vuckovic, F.~Biljecki, C.~T. Silva, M.~Lage, and N.~Ferreira.
\newblock The state of the art in visual analytics for {3D} urban data.
\newblock {\em Computer Graphics Forum}, 43(3):e15112, 2024. \href{https://doi.org/10.1111/cgf.15112}
{doi: {{%
10\hspace{.1pt}\discretionary{.}{%
}{.}\hspace{.4pt}1111\discretionary{/}{%
}{/}cgf\hspace{.1pt}\discretionary{.}{%
}{.}\hspace{.4pt}15112}}}


\bibitem{moreira_urban_2024}
G.~Moreira, M.~Hosseini, M.~N.~A. Nipu, M.~Lage, N.~Ferreira, and F.~Miranda.
\newblock The {Urban} {Toolkit}: {A} grammar-based framework for urban visual analytics.
\newblock {\em IEEE Transactions on Visualization and Computer Graphics}, 30(1):1402--1412, 2024. \href{https://doi.org/10.1109/TVCG.2023.3326598}
{doi: {{%
10\hspace{.1pt}\discretionary{.}{%
}{.}\hspace{.4pt}1109\discretionary{/}{%
}{/}TVCG\hspace{.1pt}\discretionary{.}{%
}{.}\hspace{.4pt}2023\hspace{.1pt}\discretionary{.}{%
}{.}\hspace{.4pt}3326598}}}


\bibitem{mota_comparison_2023}
R.~Mota, N.~Ferreira, J.~D. Silva, M.~Horga, M.~Lage, L.~Ceferino, U.~Alim, E.~Sharlin, and F.~Miranda.
\newblock A comparison of spatiotemporal visualizations for {3D} urban analytics.
\newblock {\em IEEE Transactions on Visualization and Computer Graphics}, 29(1):1277--1287, 2023. \href{https://doi.org/10.1109/TVCG.2022.3209474}
{doi: {{%
10\hspace{.1pt}\discretionary{.}{%
}{.}\hspace{.4pt}1109\discretionary{/}{%
}{/}TVCG\hspace{.1pt}\discretionary{.}{%
}{.}\hspace{.4pt}2022\hspace{.1pt}\discretionary{.}{%
}{.}\hspace{.4pt}3209474}}}


\bibitem{mueller_5_2018}
A.~Mueller.
\newblock 5 reasons why {Jupyter} notebooks suck.
\newblock \url{https://towardsdatascience.com/5-reasons-why-jupyter-notebooks-suck-4dc201e27086/}, 2018.
\newblock Accessed on: Jun 2024.

\bibitem{murphy_charles_2023}
D.~Murphy.
\newblock Charles {River} {Conservancy} calls for citywide shadow protection policy {Amid} {Longwood} {Place} proposal.
\newblock \url{https://thebostonsun.com/2023/01/12/charles-river-conservancy-calls-for-citywide-shadow-protection-policy-amid-longwood-place-proposal/}, 2023.
\newblock Accessed on: Jun 2024.

\bibitem{nikolopoulou_thermal_2001}
M.~Nikolopoulou, N.~Baker, and K.~Steemers.
\newblock Thermal comfort in outdoor urban spaces: {Understanding} the human parameter.
\newblock {\em Solar Energy}, 70(3):227--235, 2001. \href{https://doi.org/10.1016/S0038-092X(00)00093-1}
{doi: {{%
10\hspace{.1pt}\discretionary{.}{%
}{.}\hspace{.4pt}1016\discretionary{/}{%
}{/}S0038\discretionary{%
}{-}{-}092X\discretionary{%
}{(}{(}00\discretionary{)}{%
}{)}00093\discretionary{%
}{-}{-}1}}}


\bibitem{ogie_crowdsourced_2019}
R.~I. Ogie, R.~J. Clarke, H.~Forehead, and P.~Perez.
\newblock Crowdsourced social media data for disaster management: {Lessons} from the {PetaJakarta}.org project.
\newblock {\em Computers, Environment and Urban Systems}, 73:108--117, 2019. \href{https://doi.org/10.1016/j.compenvurbsys.2018.09.002}
{doi: {{%
10\hspace{.1pt}\discretionary{.}{%
}{.}\hspace{.4pt}1016\discretionary{/}{%
}{/}j\hspace{.1pt}\discretionary{.}{%
}{.}\hspace{.4pt}compenvurbsys\hspace{.1pt}\discretionary{.}{%
}{.}\hspace{.4pt}2018\hspace{.1pt}\discretionary{.}{%
}{.}\hspace{.4pt}09\hspace{.1pt}\discretionary{.}{%
}{.}\hspace{.4pt}002}}}


\bibitem{parker_scirun_1995}
S.~G. Parker and C.~R. Johnson.
\newblock {SCIRun}: {A} scientific programming environment for computational steering.
\newblock In {\em Proceedings of the {ACM}/{IEEE} {Conference} on {Supercomputing}}, pp. 52--es. ACM, 1995. \href{https://doi.org/10.1145/224170.224354}
{doi: {{%
10\hspace{.1pt}\discretionary{.}{%
}{.}\hspace{.4pt}1145\discretionary{/}{%
}{/}224170\hspace{.1pt}\discretionary{.}{%
}{.}\hspace{.4pt}224354}}}


\bibitem{pimentel_large-scale_2019}
J.~F. Pimentel, L.~Murta, V.~Braganholo, and J.~Freire.
\newblock A large-scale study about quality and reproducibility of {Jupyter} notebooks.
\newblock In {\em {IEEE}/{ACM} {International} {Conference} on {Mining} {Software} {Repositories}}, pp. 507--517. IEEE, 2019. \href{https://doi.org/10.1109/MSR.2019.00077}
{doi: {{%
10\hspace{.1pt}\discretionary{.}{%
}{.}\hspace{.4pt}1109\discretionary{/}{%
}{/}MSR\hspace{.1pt}\discretionary{.}{%
}{.}\hspace{.4pt}2019\hspace{.1pt}\discretionary{.}{%
}{.}\hspace{.4pt}00077}}}


\bibitem{rulff_urban_2022}
J.~Rulff, F.~Miranda, M.~Hosseini, M.~Lage, M.~Cartwright, G.~Dove, J.~Bello, and C.~T. Silva.
\newblock {Urban Rhapsody}: {Large}-scale exploration of urban soundscapes.
\newblock {\em Computer Graphics Forum}, 41(3):209--221, 2022. \href{https://doi.org/10.1111/cgf.14534}
{doi: {{%
10\hspace{.1pt}\discretionary{.}{%
}{.}\hspace{.4pt}1111\discretionary{/}{%
}{/}cgf\hspace{.1pt}\discretionary{.}{%
}{.}\hspace{.4pt}14534}}}


\bibitem{saha_visualizing_2022}
M.~Saha, S.~Patil, E.~Cho, E.~Y.-Y. Cheng, C.~Horng, D.~Chauhan, R.~Kangas, R.~McGovern, A.~Li, J.~Heer, and J.~E. Froehlich.
\newblock Visualizing urban accessibility: {Investigating} multi-stakeholder perspectives through a map-based design probe study.
\newblock In {\em Proceedings of the {CHI} {Conference} on {Human} {Factors} in {Computing} {Systems}}, pp. 1--14. ACM, 2022. \href{https://doi.org/10.1145/3491102.3517460}
{doi: {{%
10\hspace{.1pt}\discretionary{.}{%
}{.}\hspace{.4pt}1145\discretionary{/}{%
}{/}3491102\hspace{.1pt}\discretionary{.}{%
}{.}\hspace{.4pt}3517460}}}


\bibitem{saha_project_2019}
M.~Saha, M.~Saugstad, H.~T. Maddali, A.~Zeng, R.~Holland, S.~Bower, A.~Dash, S.~Chen, A.~Li, K.~Hara, and {others}.
\newblock {Project Sidewalk}: {A} web-based crowdsourcing tool for collecting sidewalk accessibility data at scale.
\newblock In {\em Proceedings of the {CHI} {Conference} on {Human} {Factors} in {Computing} {Systems}}, pp. 1--14. ACM, 2019. \href{https://doi.org/10.1145/3290605.3300292}
{doi: {{%
10\hspace{.1pt}\discretionary{.}{%
}{.}\hspace{.4pt}1145\discretionary{/}{%
}{/}3290605\hspace{.1pt}\discretionary{.}{%
}{.}\hspace{.4pt}3300292}}}


\bibitem{satyanarayan_vega-lite_2017}
A.~Satyanarayan, D.~Moritz, K.~Wongsuphasawat, and J.~Heer.
\newblock Vega-{Lite}: {A} grammar of interactive graphics.
\newblock {\em IEEE Transactions on Visualization and Computer Graphics}, 23(1):341--350, 2017. \href{https://doi.org/10.1109/TVCG.2016.2599030}
{doi: {{%
10\hspace{.1pt}\discretionary{.}{%
}{.}\hspace{.4pt}1109\discretionary{/}{%
}{/}TVCG\hspace{.1pt}\discretionary{.}{%
}{.}\hspace{.4pt}2016\hspace{.1pt}\discretionary{.}{%
}{.}\hspace{.4pt}2599030}}}


\bibitem{silva_using_2011}
C.~T. Silva, E.~Anderson, E.~Santos, and J.~Freire.
\newblock Using {VisTrails} and provenance for teaching scientific visualization.
\newblock {\em Computer Graphics Forum}, 30(1):75--84, 2011. \href{https://doi.org/10.1111/j.1467-8659.2010.01830.x}
{doi: {{%
10\hspace{.1pt}\discretionary{.}{%
}{.}\hspace{.4pt}1111\discretionary{/}{%
}{/}j\hspace{.1pt}\discretionary{.}{%
}{.}\hspace{.4pt}1467\discretionary{%
}{-}{-}8659\hspace{.1pt}\discretionary{.}{%
}{.}\hspace{.4pt}2010\hspace{.1pt}\discretionary{.}{%
}{.}\hspace{.4pt}01830\hspace{.1pt}\discretionary{.}{%
}{.}\hspace{.4pt}x}}}


\bibitem{sun_budi_2021}
X.~Sun, M.~Plaudis, and Y.~Coady.
\newblock {BUDI}: {Building} urban designs interactively. {A} spatial-based visualization and collaboration platform for urban planning.
\newblock In {\em {IEEE} {Annual} {Information} {Technology}, {Electronics} and {Mobile} {Communication} {Conference}}, pp. 888--895. IEEE, 2021. \href{https://doi.org/10.1109/IEMCON53756.2021.9623112}
{doi: {{%
10\hspace{.1pt}\discretionary{.}{%
}{.}\hspace{.4pt}1109\discretionary{/}{%
}{/}IEMCON53756\hspace{.1pt}\discretionary{.}{%
}{.}\hspace{.4pt}2021\hspace{.1pt}\discretionary{.}{%
}{.}\hspace{.4pt}9623112}}}


\bibitem{tierny_topology_2018}
J.~Tierny, G.~Favelier, J.~A. Levine, C.~Gueunet, and M.~Michaux.
\newblock The {Topology} {ToolKit}.
\newblock {\em IEEE Transactions on Visualization and Computer Graphics}, 24(1):832--842, 2018. \href{https://doi.org/10.1109/TVCG.2017.2743938}
{doi: {{%
10\hspace{.1pt}\discretionary{.}{%
}{.}\hspace{.4pt}1109\discretionary{/}{%
}{/}TVCG\hspace{.1pt}\discretionary{.}{%
}{.}\hspace{.4pt}2017\hspace{.1pt}\discretionary{.}{%
}{.}\hspace{.4pt}2743938}}}


\bibitem{wang_how_2019}
A.~Y. Wang, A.~Mittal, C.~Brooks, and S.~Oney.
\newblock How data scientists use computational notebooks for real-time collaboration.
\newblock {\em Proceedings of the ACM on Human-Computer Interaction}, 3(CSCW):1--30, 2019. \href{https://doi.org/10.1145/3359141}
{doi: {{%
10\hspace{.1pt}\discretionary{.}{%
}{.}\hspace{.4pt}1145\discretionary{/}{%
}{/}3359141}}}


\bibitem{wolf_protect_2023}
S.~Wolf.
\newblock Protect the {Emerald} {Necklace} parks - {Please} don't take our sunlight away.
\newblock \url{https://www.change.org/p/protect-the-emerald-necklace-parks-please-don-t-take-our-sunlight-away}, 2023.
\newblock Accessed on: Jun 2024.

\bibitem{wongsuphasawat_goals_2019}
K.~Wongsuphasawat, Y.~Liu, and J.~Heer.
\newblock Goals, process, and challenges of exploratory data analysis: {An} interview study.
\newblock {\em arXiv}, pp. 1--10, 2019. \href{https://doi.org/10.48550/arXiv.1911.00568}
{doi: {{%
10\hspace{.1pt}\discretionary{.}{%
}{.}\hspace{.4pt}48550\discretionary{/}{%
}{/}arXiv\hspace{.1pt}\discretionary{.}{%
}{.}\hspace{.4pt}1911\hspace{.1pt}\discretionary{.}{%
}{.}\hspace{.4pt}00568}}}


\bibitem{wu_grand_2023}
A.~Wu, D.~Deng, M.~Chen, S.~Liu, D.~Keim, R.~Maciejewski, S.~Miksch, H.~Strobelt, F.~Viégas, and M.~Wattenberg.
\newblock Grand challenges in visual analytics applications.
\newblock {\em IEEE Computer Graphics and Applications}, 43(5):83--90, 2023. \href{https://doi.org/10.1109/MCG.2023.3284620}
{doi: {{%
10\hspace{.1pt}\discretionary{.}{%
}{.}\hspace{.4pt}1109\discretionary{/}{%
}{/}MCG\hspace{.1pt}\discretionary{.}{%
}{.}\hspace{.4pt}2023\hspace{.1pt}\discretionary{.}{%
}{.}\hspace{.4pt}3284620}}}


\bibitem{wu_defence_2023}
A.~Wu, D.~Deng, F.~Cheng, Y.~Wu, S.~Liu, and H.~Qu.
\newblock In defence of visual analytics systems: {Replies} to critics.
\newblock {\em IEEE Transactions on Visualization and Computer Graphics}, 29(1):1026--1036, 2023. \href{https://doi.org/10.1109/TVCG.2022.3209360}
{doi: {{%
10\hspace{.1pt}\discretionary{.}{%
}{.}\hspace{.4pt}1109\discretionary{/}{%
}{/}TVCG\hspace{.1pt}\discretionary{.}{%
}{.}\hspace{.4pt}2022\hspace{.1pt}\discretionary{.}{%
}{.}\hspace{.4pt}3209360}}}


\bibitem{xie_first_2018}
Y.~Xie.
\newblock The first notebook war.
\newblock \url{https://yihui.org/en/2018/09/notebook-war/}, 2018.
\newblock Accessed on: Jun 2024.

\bibitem{yap_free_2022}
W.~Yap, P.~Janssen, and F.~Biljecki.
\newblock Free and open source urbanism: {Software} for urban planning practice.
\newblock {\em Computers, Environment and Urban Systems}, 96:101825, 2022. \href{https://doi.org/10.1016/j.compenvurbsys.2022.101825}
{doi: {{%
10\hspace{.1pt}\discretionary{.}{%
}{.}\hspace{.4pt}1016\discretionary{/}{%
}{/}j\hspace{.1pt}\discretionary{.}{%
}{.}\hspace{.4pt}compenvurbsys\hspace{.1pt}\discretionary{.}{%
}{.}\hspace{.4pt}2022\hspace{.1pt}\discretionary{.}{%
}{.}\hspace{.4pt}101825}}}


\bibitem{yu_visflow_2017}
B.~Yu and C.~T. Silva.
\newblock {VisFlow} - {Web}-based visualization framework for tabular data with a subset flow model.
\newblock {\em IEEE Transactions on Visualization and Computer Graphics}, 23(1):251--260, 2017. \href{https://doi.org/10.1109/TVCG.2016.2598497}
{doi: {{%
10\hspace{.1pt}\discretionary{.}{%
}{.}\hspace{.4pt}1109\discretionary{/}{%
}{/}TVCG\hspace{.1pt}\discretionary{.}{%
}{.}\hspace{.4pt}2016\hspace{.1pt}\discretionary{.}{%
}{.}\hspace{.4pt}2598497}}}


\bibitem{zheng_urban_2014}
Y.~Zheng, L.~Capra, O.~Wolfson, and H.~Yang.
\newblock Urban computing: {Concepts}, methodologies, and applications.
\newblock {\em ACM Transactions on Intelligent Systems and Technology}, 5(3):1--55, 2014. \href{https://doi.org/10.1145/2629592}
{doi: {{%
10\hspace{.1pt}\discretionary{.}{%
}{.}\hspace{.4pt}1145\discretionary{/}{%
}{/}2629592}}}


\bibitem{zheng_visual_2016}
Y.~Zheng, W.~Wu, Y.~Chen, H.~Qu, and L.~M. Ni.
\newblock Visual analytics in urban computing: {An} overview.
\newblock {\em IEEE Transactions on Big Data}, 2(3):276--296, 2016. \href{https://doi.org/10.1109/TBDATA.2016.2586447}
{doi: {{%
10\hspace{.1pt}\discretionary{.}{%
}{.}\hspace{.4pt}1109\discretionary{/}{%
}{/}TBDATA\hspace{.1pt}\discretionary{.}{%
}{.}\hspace{.4pt}2016\hspace{.1pt}\discretionary{.}{%
}{.}\hspace{.4pt}2586447}}}


\bibitem{zhu_big_2019}
L.~Zhu, F.~R. Yu, Y.~Wang, B.~Ning, and T.~Tang.
\newblock Big data analytics in intelligent transportation systems: {A} survey.
\newblock {\em IEEE Transactions on Intelligent Transportation Systems}, 20(1):383--398, 2019. \href{https://doi.org/10.1109/TITS.2018.2815678}
{doi: {{%
10\hspace{.1pt}\discretionary{.}{%
}{.}\hspace{.4pt}1109\discretionary{/}{%
}{/}TITS\hspace{.1pt}\discretionary{.}{%
}{.}\hspace{.4pt}2018\hspace{.1pt}\discretionary{.}{%
}{.}\hspace{.4pt}2815678}}}


\bibitem{ziegler_need-finding_2023}
P.~Ziegler and S.~E. Chasins.
\newblock A need-finding study with users of geospatial data.
\newblock In {\em Proceedings of the {CHI} {Conference} on {Human} {Factors} in {Computing} {Systems}}, pp. 1--16. ACM, 2023. \href{https://doi.org/10.1145/3544548.3581370}
{doi: {{%
10\hspace{.1pt}\discretionary{.}{%
}{.}\hspace{.4pt}1145\discretionary{/}{%
}{/}3544548\hspace{.1pt}\discretionary{.}{%
}{.}\hspace{.4pt}3581370}}}


\end{thebibliography}

\end{document}